\documentclass{article}[10pt]

\usepackage{natbib}

\usepackage{eucal}
\usepackage{natbib}
\usepackage{amsfonts}
\usepackage{graphicx}
\usepackage{psfrag}

\begin{document}
%-------------------------------------------------------------------
% Defines (OG)
%-------------------------------------------------------------------
%
\def\bone{{\bf 1}}
\def\bzero{{\bf 0}}
\def\DIV {\hbox{\rm div}}
\def\Grad{\hbox{\rm Grad}}
\def\sym{\mathop{\rm sym}\nolimits}
\def\dev{\mathop{\rm dev}\nolimits}
\def\Dev{\mathop {\rm DEV}\nolimits}
\def\jmpdelu{{\lbrack\!\lbrack \Delta u\rbrack\!\rbrack}}
\def\jmpudot{{\lbrack\!\lbrack\dot u\rbrack\!\rbrack}}
\def\jmpu{{\lbrack\!\lbrack u\rbrack\!\rbrack}}
\def\jmphi{{\lbrack\!\lbrack\varphi\rbrack\!\rbrack}}
\def\ljmp{{\lbrack\!\lbrack}}
\def\rjmp{{\rbrack\!\rbrack}}
\def\sB{{\mathcal B}}
\def\sU{{\mathcal U}}
\def\sC{{\mathcal C}}
\def\sS{{\mathcal S}}
\def\sV{{\mathcal V}}
\def\sL{{\mathcal L}}
\def\sO{{\mathcal O}}
\def\sH{{\mathcal H}}
\def\sG{{\mathcal G}}
\def\sM{{\mathcal M}}
\def\sN{{\mathcal N}}
\def\sF{{\mathcal F}}
\def\sW{{\mathcal W}}
\def\sT{{\mathcal T}}
\def\sR{{\mathcal R}}
\def\sJ{{\mathcal J}}
\def\sK{{\mathcal K}}
\def\sE{{\mathcal E}}
\def\sS{{\mathcal S}}
\def\sH{{\mathcal H}}
\def\sD{{\mathcal D}}
\def\sG{{\mathcal G}}
\def\sP{{\mathcal P}}
%---------------------------------------------------------
%               Bold Face Math Characters:
%               All In Format: \B***** .
%---------------------------------------------------------
\def\Bnabla{\mbox{\boldmath$\nabla$}}
\def\BGamma{\mbox{\boldmath$\Gamma$}}
\def\BDelta{\mbox{\boldmath$\Delta$}}
\def\BTheta{\mbox{\boldmath$\Theta$}}
\def\BLambda{\mbox{\boldmath$\Lambda$}}
\def\BXi{\mbox{\boldmath$\Xi$}}
\def\BPi{\mbox{\boldmath$\Pi$}}
\def\BSigma{\mbox{\boldmath$\Sigma$}}
\def\BUpsilon{\mbox{\boldmath$\Upsilon$}}
\def\BPhi{\mbox{\boldmath$\Phi$}}
\def\BPsi{\mbox{\boldmath$\Psi$}}
\def\BOmega{\mbox{\boldmath$\Omega$}}
\def\Balpha{\mbox{\boldmath$\alpha$}}
\def\Bbeta{\mbox{\boldmath$\beta$}}
\def\Bgamma{\mbox{\boldmath$\gamma$}}
\def\Bdelta{\mbox{\boldmath$\delta$}}
\def\Bepsilon{\mbox{\boldmath$\epsilon$}}
\def\Bzeta{\mbox{\boldmath$\zeta$}}
\def\Beta{\mbox{\boldmath$\eta$}}
\def\Btheta{\mbox{\boldmath$\theta$}}
\def\Biota{\mbox{\boldmath$\iota$}}
\def\Bkappa{\mbox{\boldmath$\kappa$}}
\def\Blambda{\mbox{\boldmath$\lambda$}}
\def\Bmu{\mbox{\boldmath$\mu$}}
\def\Bnu{\mbox{\boldmath$\nu$}}
\def\Bxi{\mbox{\boldmath$\xi$}}
\def\Bpi{\mbox{\boldmath$\pi$}}
\def\Brho{\mbox{\boldmath$\rho$}}
\def\Bsigma{\mbox{\boldmath$\sigma$}}
\def\Btau{\mbox{\boldmath$\tau$}}
\def\Bupsilon{\mbox{\boldmath$\upsilon$}}
\def\Bphi{\mbox{\boldmath$\phi$}}
\def\Bchi{\mbox{\boldmath$\chi$}}
\def\Bpsi{\mbox{\boldmath$\psi$}}
\def\Bomega{\mbox{\boldmath$\omega$}}
\def\Bvarepsilon{\mbox{\boldmath$\varepsilon$}}
\def\Bvartheta{\mbox{\boldmath$\vartheta$}}
\def\Bvarpi{\mbox{\boldmath$\varpi$}}
\def\Bvarrho{\mbox{\boldmath$\varrho$}}
\def\Bvarsigma{\mbox{\boldmath$\varsigma$}}
\def\Bvarphi{\mbox{\boldmath$\varphi$}}
%---------------------------------------------------------
%               Bold Face Math Italic:
%               All In Format: \b* .
%---------------------------------------------------------
\def\bA{\mbox{\boldmath$ A$}}
\def\bB{\mbox{\boldmath$ B$}}
\def\bC{\mbox{\boldmath$ C$}}
\def\bD{\mbox{\boldmath$ D$}}
\def\bE{\mbox{\boldmath$ E$}}
\def\bF{\mbox{\boldmath$ F$}}
\def\bG{\mbox{\boldmath$ G$}}
\def\bH{\mbox{\boldmath$ H$}}
\def\bI{\mbox{\boldmath$ I$}}
\def\bJ{\mbox{\boldmath$ J$}}
\def\bK{\mbox{\boldmath$ K$}}
\def\bL{\mbox{\boldmath$ L$}}
\def\bM{\mbox{\boldmath$ M$}}
\def\bN{\mbox{\boldmath$ N$}}
\def\bO{\mbox{\boldmath$ O$}}
\def\bP{\mbox{\boldmath$ P$}}
\def\bQ{\mbox{\boldmath$ Q$}}
\def\bR{\mbox{\boldmath$ R$}}
\def\bS{\mbox{\boldmath$ S$}}
\def\bT{\mbox{\boldmath$ T$}}
\def\bU{\mbox{\boldmath$ U$}}
\def\bV{\mbox{\boldmath$ V$}}
\def\bW{\mbox{\boldmath$ W$}}
\def\bX{\mbox{\boldmath$ X$}}
\def\bY{\mbox{\boldmath$ Y$}}
\def\bZ{\mbox{\boldmath$ Z$}}
\def\ba{\mbox{\boldmath$ a$}}
\def\bb{\mbox{\boldmath$ b$}}
\def\bc{\mbox{\boldmath$ c$}}
\def\bd{\mbox{\boldmath$ d$}}
\def\be{\mbox{\boldmath$ e$}}
\def\bff{\mbox{\boldmath$ f$}}
\def\bg{\mbox{\boldmath$ g$}}
\def\bh{\mbox{\boldmath$ h$}}
\def\bi{\mbox{\boldmath$ i$}}
\def\bj{\mbox{\boldmath$ j$}}
\def\bk{\mbox{\boldmath$ k$}}
\def\bl{\mbox{\boldmath$ l$}}
\def\bm{\mbox{\boldmath$ m$}}
\def\bn{\mbox{\boldmath$ n$}}
\def\bo{\mbox{\boldmath$ o$}}
\def\bp{\mbox{\boldmath$ p$}}
\def\bq{\mbox{\boldmath$ q$}}
\def\br{\mbox{\boldmath$ r$}}
\def\bs{\mbox{\boldmath$ s$}}
\def\bt{\mbox{\boldmath$ t$}}
\def\bu{\mbox{\boldmath$ u$}}
\def\bv{\mbox{\boldmath$ v$}}
\def\bw{\mbox{\boldmath$ w$}}
\def\bx{\mbox{\boldmath$ x$}}
\def\by{\mbox{\boldmath$ y$}}
\def\bz{\mbox{\boldmath$ z$}}

\title{A continuum treatment of growth in biological tissue: The coupling of mass transport and mechanics}
\author{K. Garikipati\thanks{Asst. Professor, Department of Mechanical Engineering, {\tt
krishna@umich.edu}}, E. M. Arruda\thanks{Assoc. Professor,
Department of Mechanical Engineering and Program in Macromolecular
Science and Engineering}, K. Grosh\thanks{Assoc. Professor,
Departments of Mechanical Engineering, and Biomedical
Engineering}, H. Narayanan\thanks{Graduate research assistant,
Department of Mechanical Engineering}, S. Calve\thanks{Graduate
research assistant, Program in Macromolecular Science and
Engineering}\\
University of Michigan, Ann Arbor, Michigan 48109, USA}
\date{}
\maketitle
\begin{abstract}
Growth (and resorption) of biological tissue is formulated in the
continuum setting. The treatment is macroscopic, rather than
cellular or sub-cellular. Certain assumptions that are central to
classical continuum mechanics are revisited, the theory is
reformulated, and consequences for balance laws and constitutive
relations are deduced. The treatment incorporates multiple
species. Sources and fluxes of mass, and terms for momentum and
energy transfer between species are introduced to enhance the
classical balance laws. The transported species include:
(\romannumeral 1) a fluid phase, and (\romannumeral 2) the
precursors and byproducts of the reactions that create and break
down tissue. A notable feature is that the full extent of coupling
between mass transport and mechanics emerges from the
thermodynamics. Contributions to fluxes from the concentration
gradient, chemical potential gradient, stress gradient, body force
and inertia have not emerged in a unified fashion from previous
formulations of the problem. The present work demonstrates these
effects via a physically-consistent treatment. The presence of
multiple, interacting species requires that the formulation be
consistent with mixture theory. This requirement has far-reaching
implications. A preliminary numerical example is included to
demonstrate some aspects of the coupled formulation.
\end{abstract}

\section{Background}\label{sect1}

Development of biological tissue, when described in the
biomechanics literature, is generally broken down into the
distinct processes of \emph{growth}, \emph{remodelling} and
\emph{morphogenesis}. Growth, or conversely, resorption, involves
the addition or loss of mass. Remodelling results from a change in
microstructure, which could manifest itself as an evolution of
macroscopic quantities such as state of internal stress, stiffness
or material symmetry. It also appears sometimes as fibrosis or
hypertrophy. Morphogenesis involves both growth and remodelling,
as well as more complex changes in form. A classical example of
morphogenesis is the development of an embryo from a fertilized
egg. These terms are based on the definitions developed by
\cite{Taber:95}, and will be followed in this work. In the present
communication we will focus exclusively on growth, its continuum
formulation, and the implications that the process holds for the
standard machinery of continuum mechanics. Remodelling is treated
in an accompanying paper \citep{remodelpaper}. The larger and more
complex problem of morphogenesis will not be treated in this body
of work.

The ideas here are applicable to soft (e.g., tendon, muscle) and
hard (e.g., bone) tissue. In this paper, growth of biological
tissue will be treated at a macroscopic scale. The continuum
formulation (e.g., constitutive laws) at this scale may be
motivated by cellular, sub-cellular or molecular processes.
However, we will not explicitly model processes at this fine a
scale. The formulation can be applied with a specific tissue e.g.,
muscle, as the body of interest. Our experiments, described
separately, are on self-organizing tendon \citep{Calveetal:2003}
and cardiac muscle constructs \citep{Baaretal:2003}, engineered
{\it in vitro}.

The principal notion to be borne in mind while developing a
continuum formulation for growth is that one is presented with a
system that is open with respect to mass. Scalar mass sources and
sinks, and vectorial mass fluxes must be considered. A mass source
was first introduced in the context of biological growth by
\citet{CowinHegedus:76}. The mass flux is a more recent addition
of \citet{EpsteinMaugin:2000}, who, however, did not elaborate on
the specific nature of the transported species.
\citet{KuhlSteinmann:02} also incorporated the mass flux and
specified a Fickean diffusive constitutive law for it. In their
paper the diffusing species is the material of the tissue itself.
The approach to mass transport that is followed in our paper is
outlined in the next two paragraphs.

In order to be precise about the physiological relevance of our
formulation, we have found it appropriate to adopt a different
approach from the papers in the preceding paragraph in regard to
mass transport. We do not consider mass transport of the material
making up any tissue during growth. Instead, it is the nutrients,
enzymes and amino acids necessary for growth of tissue, byproducts
of metabolic reactions, and the tissue's fluid phase
\citep{Swartzetal:99} that undergo diffusion\footnote{We use the
terms ``mass transport'' and ``diffusion'' interchangeably.} in
our treatment. There do exist certain physiological processes in
which cells or the surrounding matrix migrate within a tissue. One
such process is observed when leukocytes (white blood cells) such
as neutrophils and monocytes are signalled to pass through a
capillary wall and are induced, by specific chemical attractors,
to migrate to a site of infection. This is the process of
chemotaxis \citep{GuytonHall:1996,Vander:2003}. The migrant cells
or matrix then participate in some form of cell proliferation or
death. Fibroblasts also migrate within the extra cellular matrix
during wound healing. A third example is the migration of stem
cells to different locations during the embryonic development of
an animal. These processes involve very \emph{short range}
diffusion, and can be treated by the approach described in this
paper. We have chosen to focus upon homeostasis, defined by
\citet{Vander:2003} as ``\dots a state of reasonably stable
balance between the physiological
variables\dots''\footnote{\citet{Vander:2003} go on to say: ``This
simple definition cannot give a complete appreciation of what
homeostasis truly entails, however. There probably is no such
thing as a physiological variable that is constant over long
periods of time. In fact, some variables undergo fairly dramatic
swings about an average value during the course of a day, yet may
still be considered `in balance'. That is because homeostasis is a
\emph{dynamic process}, not a static one.''}. Since, to the best
of our knowledge, processes of the type just described are not
observed during homeostatic tissue growth, we will ignore
transport of the solid phase of the tissue.

The processes of cell proliferation and death, hypertrophy and
atrophy, are complex and involve several cascades of biochemical
reactions. We will treat them in an elementary fashion, using
source/sink terms that govern inter-conversion of species, and the
mass fluxes that supply reactants and remove byproducts. The
treatment will be mathematical; specific constituents will not be
identified with any greater detail than to say that they are
either the tissue's solid phase, the interstitial fluid phase,
precursors of the solid phase (these would include amino acids,
nutrients and enzymes), or byproducts of reactions. We will return
to explicitly incorporate biochemical and cellular processes
within our description of mass transport in a subsequent paper.

Virtually all biological tissue consists of a solid and fluid
phase and can be treated in the context of mixture theory
\citep{TruesdellToupin:60,TruesdellNoll:65,BedfordDrumheller:1983}.
When growth is of interest, additional species (reactants and
byproducts) must be considered as outlined above. The solid phase
is an anisotropic composite that is inhomogeneous at microscopic
and macroscopic scales. The fluid, being mostly water, may be
modelled as incompressible, or compressible with a very large bulk
modulus. This level of complexity will be maintained throughout
our treatment. The use of mixture theory leads to difficulties
associated with partitioning the boundary traction into portions
corresponding to each species. \cite{RajagopalWineman:1990},
suggested a resolution to this problem that holds in the case of
saturated media---a condition that is applicable to soft
biological tissue. An alternative is to apply the theory of porous
media that grew out of the classical work of Fick and Darcy in the
1800s \citep{Terzaghi:1943,deBoer:2000}. In this approach fluxes
are introduced for each species. Since a species that diffuses
must do so within some medium, one may think of the various
constituents diffusing through the solid phase\footnote{A more
sophisticated, and physiologically-correct, description is that
the interstitial fluid diffuses relative to the solid phase, while
precursors and byproducts of reactants diffuse with respect to the
fluid.}. This strategy has been adopted in the present work.

\subsection{Recent work}

In a simplification that avoided the complexity of mixture theory
or porous media, \cite{CowinHegedus:76} accounted for the fluid
phase via irreversible sources and fluxes of momentum, energy and
entropy. This approach was also followed by
\cite{EpsteinMaugin:2000} and \cite{KuhlSteinmann:02}. While the
approach followed in the present paper, i.e. derivation of a mass
balance law with mass source and flux, and postulating sources and
fluxes for momentum, energy and entropy, has been attempted
recently by \cite{EpsteinMaugin:2000}, and
\cite{KuhlSteinmann:02}, there are important differences between
those works and our paper. Epstein and Maugin conclude that the
mass flux vanishes unless the internal energy depends upon strain
gradient terms (a second-order theory). This view ignores Fickean
diffusion (where the flux is linearly dependent upon concentration
of the relevant species). Our treatment also results in the
dependence of mass flux upon strain gradient, but without the
requirement of a strain gradient dependence of the internal
energy. Instead, the dissipation inequality motivates a
constitutive relation for the mass flux of each transported
species. When properly formulated in a thermodynamic setting, the
mass flux can be constrained to depend upon the strain/stress
gradient \emph{and} the gradient in concentration (mass per unit
system volume) of the corresponding species. The latter term is
the Fickean contribution to the mass flux. The form obtained is
essentially identical to \cite{DeGrootMazur:1984}.

While modelling hard biological tissue, it is common to assume a
Fickean flux, and a mass source that depends upon the strain
energy density \citep{HarriganHamilton:1993}, and therefore upon
the strain. This introduces coupling between mass transport and
mechanics. One possible difficulty with this approach is that one
could conceive of a mass source that satisfies other requirements,
such as the dissipation inequality, but does not depend upon any
mechanical quantities. Strain- or stress-mediated mass transport
would then be absent in the boundary value problems solved with
such a formulation.

The present paper is aimed at a complete treatment of mass
transport, coupled with mechanics, for the growth problem. Initial
sections (Sections \ref{sect2}--\ref{sect3bis}) treat the balance
of mass, balance of linear and angular momenta, the forms of the
First and Second Laws for this problem, and kinematics of growth,
respectively. The Clausius-Duhem inequality and its implications
for constitutive relations are the subject of Section \ref{sect5}.
Examples are provided as appropriate to illustrate the important
results. A preliminary numerical example appears in Section
\ref{sect6}. A discussion and conclusion are provided in Section
\ref{sect7}.

\section{Balance of mass for an open system}
\label{sect2} 
\begin{figure}[ht]
\centering\psfrag{BIGX}{\small$\bX$} \psfrag{SMX}{\small$\bx$}
\psfrag{PI}{\small$\Pi^\mathrm{\iota}$}
\psfrag{PPI}{\small$\pi^\mathrm{\iota}$}
\psfrag{MN}{\small$\bN\cdot\bM^{\iota}$}
\psfrag{MMN}{\small$\bn\cdot\bm^\mathrm{\iota}$}
\psfrag{SN}{\small$\bP\bN$} \psfrag{SSN}{\small$\Bsigma\bn$}
\psfrag{PHI}{\small$\Bvarphi$} \psfrag{WO}{\small$\Omega_0$}
\psfrag{WT}{\small$\Omega_t$}
{\includegraphics[width=14cm]{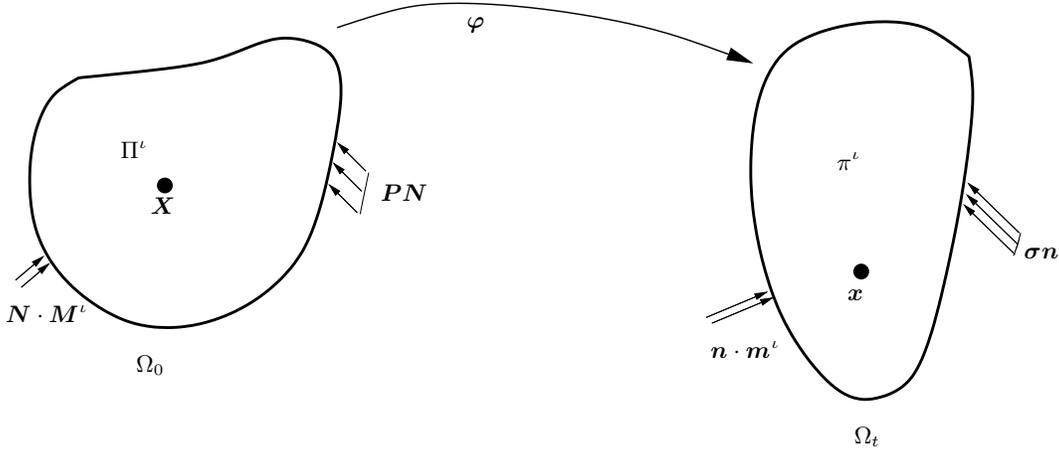}} \caption{The continuum
body with diffusing species and under stress.}\label{potato}
\end{figure}

The body of interest, $\sB$, occupies the open
region $\Omega_0\subset\mathbb{R}^3$ in the reference
configuration. Points in $\sB$ are parameterized by their
reference positions, $\bX$. The deformation of $\sB$ is a
point-to-point map, $\Bvarphi (\bX,t)\in\mathbb{R}^3$, of
$\Omega_0$, carrying the point at $\bX$ to its current position
$\bx = \Bvarphi(\bX,t)$, at time $t\in [0,T]$. In its current
configuration, $\sB$ occupies the open region $\Omega_t =
\Bvarphi_t(\Omega_0),\; \Omega_t\subset\mathbb{R}^3$ (see Figure
\ref{potato}). The tangent map of $\Bvarphi$ is the deformation
gradient $\bF :=
\partial\Bvarphi/\partial\bX$.

The body consists of several species, of which the solid phase of
the tissue is denoted $\mathrm{s}$, and the fluid phase is
$\mathrm{f}$. The remaining species, $\alpha,\dots,\omega$, are
precursors of the tissue or byproducts of its breakdown in
chemical reactions. The index $\iota$ will be used to indicate an
arbitrary species. (Where appropriate, we will use the term
``system'' to refer to the species collectively. Where the
presence of several species is an unimportant fact, we will use
the term ``body''. ) The system is open with respect to mass.
Species $\alpha,\dots,\omega$ have sources/sinks,
$\Pi^\alpha,\dots,\Pi^\omega$, and mass fluxes,
$\bM^\alpha,\dots,\bM^\omega$, respectively. The sources specify
mass production rates per unit volume of the body in its reference
configuration, $\Omega_0$. The fluxes specify mass flow rates per
unit cross-sectional area in $\Omega_0$. Importantly, \emph{these
flux vectors are defined relative to the solid phase}. As
discussed in Section \ref{sect1}, the solid tissue phase has only
a mass source/sink associated with it, and no flux. The fluid
tissue phase has only a mass flux, and no
source/sink\footnote{Considering the case of the lymphatic fluid,
this implies that lymph glands are assumed not to be present.}.

Since the solid phase, $\mathrm{s}$, does not undergo transport,
its motion is specified entirely by $\Bvarphi(\bX,t)$. We describe
the remaining species $\mathrm{f},\alpha,\dots,\omega$ as
convecting with the solid phase and diffusing with respect to it.
They therefore have a velocity relative to $\mathrm{s}$. Since the
remaining species convect with $\mathrm{s}$, it implies a local
homogenization of deformation. The modelling assumption is made
that at each point, $\bX$, the individual phases undergo the same
deformation.

We define concentrations of the species
$\rho_0^\iota=\bar{\rho}_0^\iota f^\iota$ as masses per unit
volume in $\Omega_0$. The intrinsic species density is
$\bar{\rho}_0^\iota$, and $f^\iota$ is the volume fraction of
$\iota$, for $\iota = \mathrm{s,f},\alpha,\dots,\omega$. In an
experiment it is far easier to measure the concentration,
$\rho_0^\iota$, rather than the intrinsic species density,
$\bar{\rho}_0^\iota$\footnote{In our experiments we have measured
the mass concentration $\rho_0^\iota$ of collagen in engineered
tendons grown \emph{in vitro}. These results will be presented
elsewhere \citep{Calveetal:2004}.}. The concentrations also have
the property $\sum\limits_{\iota}\rho_0^\iota = \rho_0$, the total
material density of the tissue, with the sum being over all
species $\mathrm{s,f},\alpha,\dots,\omega$. The concentrations,
$\rho_0^\iota$, change as a result of mass transport and
inter-conversion of species, implying that the total density in
the reference configuration, $\rho_0$, changes with time. They are
parameterized as $\rho_0^\iota(\bX,t)$.

\subsection{Balance of mass in the reference configuration}
\label{sect2.1} Recall that $\Omega_0$ is a fixed volume. The
statement of balance of mass of the solid phase of the tissue,
written in integral form over $\Omega_0$, is

\begin{equation}
\frac{\mathrm{d}}{\mathrm{d}t} \int\limits_{\Omega_0}
\rho_0^\mathrm{s} (\bX,t)\mathrm{d}V = \int\limits_{\Omega_0}
\Pi^\mathrm{s} (\bX,t)\mathrm{d}V. \label{massbalintA}
\end{equation}

\noindent Since the solid phase of the tissue does not undergo
mass transport, there is no associated flux. Localizing the result
gives

\begin{equation}
\frac{\partial\rho_0^\mathrm{s}}{\partial t} = \Pi^\mathrm{s},
\label{massballocA}
\end{equation}
\noindent where the explicit dependence upon position and time has
been suppressed.

The fluid phase of the tissue, $\mathrm{f}$, may be thought of as
the interstitial or lymphatic fluid that perfuses the tissue. As
explained above, we do not consider sources of fluid in the region
of interest. The fluid therefore enters and leaves $\Omega_0$ as a
flux, $\bM^\mathrm{f}$. The balance of mass in integral form is
\begin{equation}
\frac{\mathrm{d}}{\mathrm{d}t} \int\limits_{\Omega_0}
\rho_0^\mathrm{f} (\bX,t)\mathrm{d}V =
-\int_{\partial\Omega_0}\bM^\mathrm{f}(\bX,t)\cdot\bN \mathrm{d}A,
\label{massbalintB}
\end{equation}

\noindent where $\bN$ is the unit outward normal to the boundary,
$\partial\Omega_0$. Applying the Divergence Theorem to the surface
integral and localizing the result gives
\begin{equation}
\frac{\partial\rho_0^\mathrm{f}}{\partial t} = -
\Bnabla\cdot\bM^\mathrm{f}, \label{massballocB}
\end{equation}

\noindent where $\Bnabla(\bullet)$ is the gradient operator
defined on $\Omega_0$, and $\Bnabla\cdot(\bullet)$ denotes the
divergence of a vector or tensor argument on $\Omega_0$.

For the precursor and byproduct species,
$\iota=\alpha,\dots,\omega$, the balance of mass in integral form
is
\begin{equation}
\frac{\mathrm{d}}{\mathrm{d}t} \int\limits_{\Omega_0} \rho_0^\iota
(\bX,t)\mathrm{d}V = \int\limits_{\Omega_0} \Pi^\iota
(\bX,t)\mathrm{d}V
-\int_{\partial\Omega_0}\bM^\iota(\bX,t)\cdot\bN \mathrm{d}A.
\label{massbalintI}
\end{equation}

\noindent In local form it is
\begin{equation}
\frac{\partial\rho_0^\iota}{\partial t} = \Pi^\iota -
\Bnabla\cdot\bM^\iota,\;\forall\,\iota=\alpha,\dots,\omega.
\label{massballocI}
\end{equation}

\noindent Of course, this last equation is the general form of
mass balance for any species $\iota$, recalling that in
particular, $\bM^\mathrm{s} = \bzero$ and $\Pi^\mathrm{f} = 0$.
This form will be used in the development that follows.

The fluxes,
$\bM^\iota,\;\forall\,\iota=\mathrm{f},\alpha,\dots,\omega$,
represent mass transport of the fluid, of precursors to the
reaction site, and of byproducts from sites of tissue breakdown.
The sources,
$\Pi^\iota,\;\forall\iota=\mathrm{s},\alpha,\dots,\omega$, arise
from inter-conversion of species. The sources/sinks in
(\ref{massballocA}) and (\ref{massballocI}) are therefore related,
as tissue and byproducts are formed by consuming precursors (amino
acids and nutrients, for instance). To maintain a degree of
simplicity in this initial exposition, we will restrict our
description of tissue breakdown to the reverse of this reaction.

The sources, $\Pi^\iota$ for various species, satisfy a relation
that is arrived as follows: Summing (\ref{massballocI}) over all
species leads to the law of mass balance for the system,
\begin{equation}
\sum\limits_\iota\frac{\partial\rho_0^\iota}{\partial t} =
\sum\limits_\iota\left(\Pi^\iota - \Bnabla\cdot\bM^\iota\right),
\label{massballocItot}
\end{equation}

\noindent where the sum runs over all species, with
$\Pi^\mathrm{f} = 0$ and $\bM^\mathrm{s} = \bzero$. Alternatively,
in writing the mass balance equation for the system, the
interconversion terms (sources/sinks) play no role, and only the
fluxes at the boundaries need be accounted for. In integral form
we have
\begin{displaymath}
\frac{\mathrm{d}}{\mathrm{d}t}\sum\limits_\iota\int\limits_{\Omega_0}\rho_0^\iota
\mathrm{d}V =
-\sum\limits_\iota\int\limits_{\partial\Omega_0}\bM^\iota\cdot\bN
\mathrm{d}A.
\end{displaymath}

\noindent Applying the Divergence Theorem and localizing leads to
\begin{equation}
\sum\limits_\iota \frac{\partial\rho_0^\iota}{\partial t} =
\sum\limits_\iota(-\Bnabla\bM^\iota). \label{massballoctot}
\end{equation}

\noindent Comparing the equivalent forms (\ref{massballocItot})
and (\ref{massballoctot}) it emerges that the sources and sinks
satisfy
\begin{equation}
\sum\limits_\iota\Pi^\iota = 0, \label{sourcebalance}
\end{equation}

\noindent a conclusion that is consistent with classical mixture
theory \citep{TruesdellNoll:65}. The section that follows contains
an example in which the Law of Mass Action is invoked to describe
a set of inter-related sources, $\Pi^\iota$.

\subsubsection{Sources, sinks and stoichiometry: An example based upon the Law of Mass Action}\label{sect2.1.1}

The conversion of precursors to tissue and the reverse process of
its breakdown are governed by a series of chemical reactions. The
stoichiometry of these reactions varies in a limited range.
Continuing in the simple vein adopted above, it is assumed that
the formation of tissue and byproducts from precursors, and the
breakdown of tissue, are governed by the forward and reverse
directions of a single reaction:
\begin{equation}
\sum\limits_{\iota=\alpha}^{\omega} n_\iota[\iota] \longrightarrow
[\mathrm{s}]. \label{chemreac}
\end{equation}

\noindent Here, $n_\iota$ is the (possibly fractional) number of
moles of species $\iota$ in the reaction. For a tissue precursor,
$n_\iota > 0$, and for a byproduct, $n_\iota < 0$. By the Law of
Mass Action for this reaction, the rate of the forward reaction
(number of moles of $\mathrm{s}$ produced per unit time, per unit
volume in $\Omega_0$) is
$k_\mathrm{f}\prod\limits_{\iota=\alpha}^\omega
[\rho_0^\iota]^{n_\iota}$, where $\prod$ on the right hand-side
denotes a product, not to be confused with the source, $\Pi$. The
rate of the reverse reaction (number of moles of $\mathrm{s}$
consumed per unit time, per unit volume in $\Omega_0$) is
$k_\mathrm{r}[\rho_0^\mathrm{s}]$, where $k_f$ and $k_r$ are the
corresponding reaction rates. Assuming, for the purpose of this
example, that the solid phase is a single compound, let the
molecular weight of $\mathrm{s}$ be $\sM_\mathrm{s}$. From the
above arguments the source term for $\mathrm{s}$ is
\begin{equation}
\Pi^\mathrm{s} =
\left(k_\mathrm{f}\prod\limits_{\iota=\alpha}^\omega
[\rho_0^\iota]^{n_\iota} -
k_\mathrm{r}[\rho_0^\mathrm{s}]\right)\sM_\mathrm{s},
\label{sourceA}
\end{equation}

\noindent Since the formation of one mole of $\mathrm{s}$ requires
consumption of $n_\iota$ moles of $\iota$, we have
\begin{equation}
\Pi^\iota =
-\left(k_\mathrm{f}\prod\limits_{\vartheta=\alpha}^\omega
[\rho_0^\vartheta]^{n_\vartheta} -
k_\mathrm{r}[\mathrm{s}]\right)n_\iota\sM_\iota, \label{sourceI}
\end{equation}

\noindent where $\sM_\iota$ is the molecular weight of species
$\iota$. Since, due to conservation of mass, $\sM_\mathrm{s} =
\sum\limits_{\iota=\alpha}^\omega n^\iota\sM_\iota$ the sources
satisfy $\sum\limits_{\iota=\mathrm{s},\alpha}^\omega \Pi^\iota =
0$.

\subsection{Balance of mass in the current configuration}
\label{sect2.2} In the current configuration, $\Omega_t$, the
concentration, source and mass flux of species $\iota$ are
$\rho^\iota(\bx,t),\,\pi^\iota\mathrm(\bx,t)$, and
$\bm^\iota(\bx,t)$ respectively. The boundary is
$\partial\Omega_t$, and has outward normal $\bn$ (Figure
\ref{potato}). Since the deformation, $\Bvarphi(\bX,t)$, is
applied to the body (the system), standard arguments yield
$\rho^\iota = \rho_0^\iota(\mathrm{det}\bF)^{-1}$, and $\pi^\iota
= \Pi^\iota(\mathrm{det}\bF)^{-1}$. By Nanson's formula, the
normals satisfy $\bn\mathrm{d}a =
\bF^{-\mathrm{T}}\bN\mathrm{d}A$, where $\mathrm{d}A$ and
$\mathrm{d}a$ are the area elements on the boundaries
$\partial\Omega_0$ and $\partial\Omega_t$ respectively. The Piola
transform then gives $\bm^\iota =
(\mathrm{det}\bF)^{-1}\bF\bM^\iota$.

The balance of mass in integral form for the solid tissue phase is
\begin{equation}
\frac{\mathrm{d}}{\mathrm{d}t}
\int\limits_{\Omega_t}\rho^\mathrm{s}(\bx,t)\mathrm{d}v =
\int\limits_{\Omega_t}\pi^\mathrm{s}(\bx,t)\mathrm{d}v.
\label{massbalcurr1}
\end{equation}
\noindent Applying Reynolds' Transport Theorem to the left
hand-side, localizing the result and employing the product rule
gives,
\begin{equation}
\frac{\partial\rho^\mathrm{s}}{\partial t} +
\bv\cdot\Bnabla_x\rho^\mathrm{s} + \rho^\mathrm{s}
\Bnabla_x\cdot\bv = \pi^\mathrm{s},\nonumber
\end{equation}
\noindent or,
\begin{equation}
\frac{\mathrm{d}\rho^\mathrm{s}}{\mathrm{d}t} = \pi^\mathrm{s} -
\rho^\mathrm{s} \Bnabla_x\cdot\bv, \label{massbalcurr2}
\end{equation}
\noindent where $\bv(\bx,t) = \bV(\bX,t)\circ\Bvarphi^{-1}(\bx,t)$
is the spatial velocity, $\Bnabla_x(\bullet)$ is the spatial
gradient operator, and $\Bnabla_x\cdot(\bullet)$ is the spatial
divergence. The time derivative on the left hand-side in
(\ref{massbalcurr2}) is the material time derivative, that may be
written explicitly as $(\partial\rho^\mathrm{s}(\bx,t)/\partial
t)_X$, implying that the reference position is held fixed.

For the fluid tissue phase, $\mathrm{f}$, the integral form is,
\begin{eqnarray}
\frac{\mathrm{d}}{\mathrm{d}t}
\int\limits_{\Omega_t}\rho^\mathrm{f}(\bx,t)\mathrm{d}v & = & -
\int\limits_{\partial\Omega_t}\bm^\mathrm{f}(\bx,t)\cdot\bn\mathrm{d}a.
\label{massbalcurr1B}
\end{eqnarray}

\noindent Invoking the Divergence Theorem in addition to the
arguments used above for the solid phase now gives
\begin{equation}
\frac{\mathrm{d}\rho^\mathrm{f}}{\mathrm{d}t} = -
\Bnabla_x\cdot\bm^\mathrm{f} - \rho^\mathrm{f} \Bnabla_x\cdot\bv.
\label{massbalcurr2B}
\end{equation}

Finally, for the precursor and byproduct species, $\iota =
\alpha,\dots,\omega$, proceeding as above gives the local form in
$\Omega_t$:
\begin{equation}
\frac{\mathrm{d}\rho^\iota}{\mathrm{d}t} = \pi^\iota-
\Bnabla_x\cdot\bm^\iota - \rho^\iota
\Bnabla_x\cdot\bv,\;\forall\,\iota = \alpha,\dots,\omega.
\label{massbalcurr2I}
\end{equation}

\section{Balance of linear and angular momenta}
\label{sect3}

\subsection{Balance of linear momentum}
\label{sect3.1} We first write the balance of linear momentum in
the reference configuration, $\Omega_0$. The body, $\sB$, is
subject to surface traction, $\bT$, and body force per unit mass,
$\bg$. The natural boundary condition then implies that $\bT =
\sum_{\iota}\bP^\iota\bN$ on $\partial\Omega_0$, where $\bP^\iota$
is the partial first Piola-Kirchhoff stress tensor corresponding
to species $\iota$, and the index runs over all species. Thus,
$\bP^\iota\bN$ is the corresponding partial traction. The mass
fluxes, $\bM^\iota,\;(\iota = \mathrm{f},\alpha,\dots,\omega)$,
and mass sources, $\Pi^\iota,\;(\iota =
\mathrm{s},\alpha,\dots,\omega)$ make important contributions to
the balance of linear momentum, as shown below.

The body undergoes deformation, $\Bvarphi(\bX,t)$, and has a
material velocity field $\bV(\bX,t) =
\partial\Bvarphi(\bX,t)/\partial t$. In discussing momentum and energy, it proves convenient to define a
material velocity of species $\iota$ relative to the solid phase
as $\bV^\iota = (1/\rho_0^\iota)\bF\bM^\iota$. Recall (from
Section \ref{sect2}) that the remaining species are described as
deforming with the solid phase and diffusing relative to it.
Therefore $\bF$ is common to all species. The spatial velocity
corresponding to $\bV^\iota$ is $\bv^\iota =
(1/\rho^\iota)\bm^\iota = \bV^\iota$, by the Piola transform.
Since fluxes are defined relative to the solid tissue phase, which
does not diffuse, the total material velocity of the solid phase
is $\bV$, and for each of the remaining species it is $\bV +
\bV^\iota$, $\iota = \mathrm{f},\alpha,\dots,\omega$. Formally, we
can write the material velocity as  $\bV + \bV^\iota$, $\iota =
\mathrm{s},\mathrm{f},\alpha,\dots,\omega$ with the understanding
that $\bV^\mathrm{s} = \bzero$. Likewise, $\Pi^\mathrm{f} = 0$.
This convention has been adopted in the remainder of the paper.

The balance of linear momentum of species $\iota$ written in
integral form over $\Omega_0$ is,
\begin{eqnarray}
\frac{\mathrm{d}}{\mathrm{d}t} \int\limits_{\Omega_0}
\rho_0^\iota(\bV+\bV^\iota) \mathrm{d}V &=& \int\limits_{\Omega_0}
\rho^\iota_0\bg \mathrm{d}V + \int\limits_{\Omega_0}
\rho^\iota_0\bq^\iota \mathrm{d}V + \int\limits_{\Omega_0}
\Pi^\iota(\bV+\bV^\iota) \mathrm{d}V \nonumber\\
& &+ \int\limits_{\partial \Omega_0}\bP^\iota\bN \mathrm{d}A -
\int\limits_{\partial \Omega_0}(\bV+\bV^\iota)\bM^\iota\cdot\bN
\mathrm{d}A, \label{linmombalI1}
\end{eqnarray}
\noindent where $\bg$ is the body force per unit mass, and
$\bq^\iota$ is the force per unit mass exerted upon $\iota$ by the
other species present. Attention is drawn to the fact that the
mass source distributed through the volume, and the influx over
the boundary affect the rate of change of momentum in
(\ref{linmombalI1}). Summing over all species, the balance of
linear momentum for the system is obtained:
\begin{eqnarray}
\sum\limits_{\iota}\frac{\mathrm{d}}{\mathrm{d}t}
\int\limits_{\Omega_0} \rho_0^\iota(\bV+\bV^\iota) \mathrm{d}V &=&
\sum\limits_{\iota}\int\limits_{\Omega_0} \rho^\iota_0\bg
\mathrm{d}V + \sum\limits_{\iota}\int\limits_{\Omega_0}
\rho^\iota_0\bq^\iota \mathrm{d}V\nonumber\\& & +
\sum\limits_{\iota}\int\limits_{\Omega_0} \Pi^\iota(\bV+\bV^\iota)
\mathrm{d}V + \sum\limits_{\iota} \int\limits_{\partial
\Omega_0}\bP^\iota\bN \mathrm{d}A\nonumber\\
& & - \sum\limits_{\iota}\int\limits_{\partial
\Omega_0}(\bV+\bV^\iota)\bM^\iota\cdot\bN \mathrm{d}A,
\label{linmombalall}
\end{eqnarray}

The interaction forces, $\rho_0^\iota\bq^\iota$, satisfy a
relation with the mass sources, $\Pi^\iota$, that is elucidated by
the following argument: The rate of change of momentum of the
entire system is affected by external agents only, and is
independent of internal interactions of any nature ($\bq^\iota$
and $\Pi^\iota$). This observation leads to the following
equivalent expression for the rate of change of linear momentum of
the system:

\begin{eqnarray}
\sum\limits_{\iota}\frac{\mathrm{d}}{\mathrm{d}t}
\int\limits_{\Omega_0} \rho_0^\iota(\bV+\bV^\iota) \mathrm{d}V &=&
\int\limits_{\Omega_0}\rho_0\bg \mathrm{d}V +
\int\limits_{\partial \Omega_0}\bP\bN \mathrm{d}A \nonumber\\
& &- \sum\limits_{\iota}\int\limits_{\partial
\Omega_0}(\bV+\bV^\iota)\bM^\iota\cdot\bN \mathrm{d}A.
\label{linmombalsys}
\end{eqnarray}

\noindent Here, $\bP = \sum\limits_{\iota}\bP^\iota$ and
$\rho_0=\sum\limits_{\iota}\rho_0^\iota$. Since both
(\ref{linmombalall}) and (\ref{linmombalsys}) represent the
balance of linear momentum of the system, it follows that,

\begin{eqnarray}
\sum\limits_{\iota}\int\limits_{\Omega_0} \rho^\iota_0\bq^\iota
\mathrm{d}V + \sum\limits_{\iota}\int\limits_{\Omega_0}
\Pi^\iota(\bV+\bV^\iota) \mathrm{d}V = 0
\end{eqnarray}

Recalling the relation between the sources (\ref{sourcebalance}),
and localizing leads to

\begin{eqnarray}
\sum\limits_{\iota}\left(\rho^\iota_0\bq^\iota +
\Pi^\iota\bV^\iota\right)=0, \label{interforcebalance}
\end{eqnarray}

\noindent a result that is also consistent with classical mixture
theory \citep{TruesdellNoll:65}.

Having established (\ref{interforcebalance}) we return to the
balance of linear momentum for a single species
(\ref{linmombalI1}) in order to simplify it. Writing
$(\bV+\bV^\iota)[\bM^\iota\cdot\bN]$ as
$((\bV+\bV^\iota)\otimes\bM^\iota)\bN$, and using the Divergence
Theorem,
\begin{eqnarray}
\int\limits_{\Omega_0} \left(\frac{\partial\rho_0^\iota}{\partial
t}\left(\bV+\bV^\iota\right) +
\rho_0^\iota\frac{\partial}{\partial
t}\left(\bV+\bV^\iota\right)\right) \mathrm{d}V =
\int\limits_{\Omega_0}\rho^\iota_0\left(\bg+\bq^\iota\right)\mathrm{d}V\qquad& &\nonumber\\
+\int\limits_{\Omega_0}\left(\Pi^\iota\left(\bV+\bV^\iota\right) +
\Bnabla\cdot\bP^\iota\right)
\mathrm{d}V& & \nonumber\\
-
\int\limits_{\Omega_0}\Bnabla\cdot\left(\left(\bV+\bV^\iota\right)\otimes\bM^\iota\right)\mathrm{d}V
& &
\end{eqnarray}

\noindent Using the mass balance equation (\ref{massballocI}), and
applying the product rule to the last term gives
\begin{eqnarray}
\int\limits_{\Omega_0} \rho_0^\iota\frac{\partial}{\partial
t}\left(\bV+\bV^\iota\right) \mathrm{d}V &=&
\int\limits_{\Omega_0}\rho^\iota_0\left(\bg+\bq^\iota\right)\mathrm{d}V\nonumber\\
& &+
\int\limits_{\Omega_0}\left(\Bnabla\cdot\bP^\iota-\left(\Bnabla\left(\bV+\bV^\iota\right)\right)\bM^\iota\right)\mathrm{d}V
\end{eqnarray}

\noindent Localizing this result gives the balance of linear
momentum for a single species in the reference configuration:

\begin{equation}
\rho_0^\iota\frac{\partial}{\partial t}\left(\bV+\bV^\iota\right)
= \rho^\iota_0\left(\bg+\bq^\iota\right) +
\Bnabla\cdot\bP^\iota-\left(\Bnabla\left(\bV+\bV^\iota\right)\right)\bM^\iota
\label{ballinmomrefI}
\end{equation}

The balance of linear momentum for a single species in the current
configuration, $\Omega_t$, is obtained via similar arguments and
the Reynolds Transport Theorem:
\begin{eqnarray}
\rho^\iota\frac{\partial}{\partial t}\left(\bv+\bv^\iota\right)
&=& \rho^\iota\left(\bg+\bq^\iota\right) +
\Bnabla_x\cdot\Bsigma^\iota\nonumber\\
& & - \left(\Bnabla_x\left(\bv+\bv^\iota\right)\right)\bm^\iota -
\rho^\iota\left(\Bnabla_x\left(\bv+\bv^\iota\right)\right)\bv,
\label{ballinmomcurrI}
\end{eqnarray}

\noindent where $\Bsigma^\iota =
(\mathrm{det}\bF)^{-1}\bP^\iota\bF^\mathrm{T}$ is the partial
Cauchy stress of species $\iota$.

\subsection{Angular Momentum}
\label{sect3.2} For the purely mechanical theory, balance of
angular momentum implies that the Cauchy stress is symmetric:
$\Bsigma^\iota = \Bsigma^{\iota^\mathrm{T}}$. We now re-examine
this result in the presence of mass transport. At the outset one
might expect a statement on the symmetry of some stress or
stress-like quantity. We derive this result for any species,
$\iota$, beginning with the integral form of balance of angular
momentum written over $\Omega_0$.

\begin{eqnarray}
\frac{\mathrm{d}}{\mathrm{d}t} \int\limits_{\Omega_0} \Bvarphi
\times \rho^\iota_0(\bV+\bV^\iota)\mathrm{d}V &=&
\int\limits_{\Omega_0}\Bvarphi\times\left[\rho^\iota_0\left(\bg+\bq^\iota\right)+\Pi^\iota\left(\bV+\bV^\iota\right)\right]\mathrm{d}V\nonumber\\
& &+
\int\limits_{\partial\Omega_0}\Bvarphi\times\left(\bP^\iota-\left(\bV+\bV^\iota\right)\otimes\bM^\iota\right)\bN
\mathrm{d}A
\end{eqnarray}

\noindent Applying properties of the cross product, the Divergence
Theorem and product rule gives
\begin{eqnarray}
\int\limits_{\Omega_0}\bV\times\rho^\iota_0\bV^\iota +
\Bvarphi\times \left(\frac{\partial\rho^\iota_0}{\partial
t}\left(\bV+\bV^\iota\right)+ \rho^\iota_0\frac{\partial}{\partial
t}\left(\bV+\bV^\iota\right)\right)\mathrm{d}V =\qquad\quad& &\nonumber\\
\int\limits_{\Omega_0}\Bvarphi\times\rho^\iota_0\left(\bg+\bq^\iota+\Pi^\iota\left(\bV+\bV^\iota\right)\right)\mathrm{d}V& &\nonumber\\
 +
\int\limits_{\Omega_0}\left(\Bvarphi\times\Bnabla\cdot\bP^\iota-\Bvarphi\times\left(\Bnabla\left(\bV+\bV^\iota\right)\bM^\iota\right)\right)\mathrm{d}V
& &\nonumber\\
\int\limits_{\Omega_0}\left(-\Bvarphi\times\left(\bV+\bV^\iota\right)\Bnabla\cdot\left(\bM^\iota\right)\right)
\mathrm{d}V & &\nonumber\\
-
\int\limits_{\Omega_0}\Bepsilon\colon\left(\left(\bP^\iota-\left(\bV+\bV^\iota\right)\otimes\bM^\iota\right)\bF^\mathrm{T}\right)\mathrm{d}V,
\end{eqnarray}

\noindent where $\Bepsilon$ is the permutation symbol, and
$\Bepsilon\colon\bA$ is written as $\epsilon_{ijk}A_{jk}$ in
indicial form, for any second-order tensor $\bA$. Using the mass
balance equation (\ref{massballocI}), and balance of linear
momentum (\ref{ballinmomrefI}), we have
\begin{displaymath}
\int\limits_{\Omega_0}\bV\times\rho^\iota_0\bV^\iota\mathrm{d}V =
-\int\limits_{\Omega_0}\Bepsilon\colon\left(\left(\bP^\iota-\left(\bV+\bV^\iota\right)\otimes\underbrace{\bM^\iota}_{\rho_0^\iota\bF^{-1}\bV^\iota}\right)\bF^\mathrm{T}\right)\mathrm{d}V.
\end{displaymath}

\noindent Recalling the relation of the permutation symbol to the
cross product, and the indicated relation between $\bM^\iota$ and
$\bV^\iota$ leads to
\begin{equation}
\bzero =
-\int\limits_{\Omega_0}\Bepsilon\colon\left(\left(\bP^\iota-\bV^\iota\otimes\rho_0^\iota\bF^{-1}\bV^\iota\right)\bF^\mathrm{T}\right)\mathrm{d}V.
\end{equation}

\noindent Localizing this result and again applying the properties
of the permutation symbol we are led to the symmetry condition,
\begin{equation}
\left(\bP^\iota-\bV^\iota\otimes\rho^\iota_0\bF^{-1}\bV^\iota\right)\bF^\mathrm{T}
=
\bF\left(\bP^\iota-\bV^\iota\otimes\rho^\iota_0\bF^{-1}\bV^\iota\right)^\mathrm{T}.
\end{equation}

\noindent But, $(\bV^\iota\otimes\bF^{-1}\bV^\iota)\bF^\mathrm{T}
= \bV^\iota\otimes\bV^\iota$. Thus, the symmetry
$\bP^\iota\bF^\mathrm{T} = \bF(\bP^\iota)^\mathrm{T}$ that results
from conservation of angular momentum for the purely mechanical
theory, is retained in this case. The partial Cauchy stresses are
therefore symmetric: $\Bsigma^\iota = \Bsigma^{\iota^\mathrm{T}}$.
This is in contrast with the non-symmetric Cauchy stress arrived
at by \citet{EpsteinMaugin:2000}. The origin of this difference
lies in the fact that these authors use a single species with $\bV
=
\partial\Bvarphi/\partial t$ as the material velocity, rather than
multiple species with material velocities $\bV + \bV^\iota$.

\section{Balance of energy and the entropy inequality}
\label{sect4}

\subsection{Balance of energy}\label{sect4.1}
Since mass is undergoing transport with respect to $\sB$, and
inter-conversion between species $\iota =
\mathrm{s},\alpha,\dots,\omega$, it is appropriate to work with
energy and energy-like quantities per unit mass. In addition to
the terms introduced in previous sections, the internal energy per
unit mass of species $\iota$ is denoted $e^\iota$; the heat supply
to species $\iota$ per unit mass of that species is $r^\iota$; and
the partial heat flux vector of $\iota$ is $\bQ^\iota$, defined on
$\Omega_0$. An interaction energy appears between species: The
energy transferred to $\iota$ by all other species is
$\tilde{e}^\iota$, per unit mass of $\iota$. In the arguments to
follow in this section we will use the fluxes $\bM^\iota$
\emph{and} the associated velocities, $\bV^\iota$. Working in
$\Omega_0$, we relate the rate of change of internal and kinetic
energies of species $\iota$ to the work done on $\iota$ by
mechanical loads, processes of mass production and transport,
heating and energy transfer:

\begin{eqnarray}
\frac{\mathrm{d}}{\mathrm{d}t} \int\limits_{\Omega_0}\rho_0^\iota
\left ( e^\iota + \frac{1}{2} \Vert\bV+\bV^\iota\Vert^2 \right )
\mathrm{d}V =  \int\limits_{\Omega_0} \left(\rho_0^\iota\bg
\cdot\left(\bV+\bV^\iota\right) + \rho_0^\iota r^\iota
\right)\mathrm{d}V\qquad& &\nonumber\\
+\int\limits_{\Omega_0}\rho_0^\iota\bq^\iota\cdot(\bV+\bV^\iota)\mathrm{d}V&
&\nonumber\\
+ \int\limits_{\Omega_0} \left(\Pi^\iota\left(e^\iota
+ \frac{1}{2}\Vert\bV+\bV^\iota\Vert^2\right)+\rho^\iota_0\tilde{e}^\iota\right)\mathrm{d}V & & \nonumber \\
+\int\limits_{\partial
\Omega_0}\left(\left(\bV+\bV^\iota\right)\cdot\bP^\iota -
\bM^\iota\left(e^\iota +\frac{1}{2}
\Vert\bV+\bV^\iota\Vert^2\right) -
\bQ^\iota\right)\cdot\bN\mathrm{d}A. \label{energyspecies1}
\end{eqnarray}

\noindent Summing over all species, the rate of change of energy
of the system is,

\begin{eqnarray}
\sum\limits_{\iota}\frac{\mathrm{d}}{\mathrm{d}t}
\int\limits_{\Omega_0}\rho_0^\iota \left ( e^\iota + \frac{1}{2}
\Vert\bV+\bV^\iota\Vert^2 \right ) \mathrm{d}V =
\sum\limits_{\iota}\int\limits_{\Omega_0} \left(\rho_0^\iota\bg
\cdot\left(\bV+\bV^\iota\right) + \rho_0^\iota r^\iota \right)\mathrm{d}V& &\nonumber\\
+\sum\limits_{\iota}\int\limits_{\Omega_0}\rho_0^\iota\bq^\iota\cdot(\bV+\bV^\iota)\mathrm{d}V&
&\nonumber\\
+ \sum\limits_{\iota}\int\limits_{\Omega_0}
\left(\Pi^\iota\left(e^\iota
+ \frac{1}{2}\Vert\bV+\bV^\iota\Vert^2\right)+\rho^\iota_0\tilde{e}^\iota\right)\mathrm{d}V & & \nonumber \\
+\sum\limits_{\iota}\int\limits_{\partial
\Omega_0}\left(\left(\bV+\bV^\iota\right)\cdot\bP^\iota -
\bM^\iota\left(e^\iota +\frac{1}{2}
\Vert\bV+\bV^\iota\Vert^2\right) -
\bQ^\iota\right)\cdot\bN\mathrm{d}A.\quad& & \label{energysum}
\end{eqnarray}

The inter-species energy transfers are related to interaction
forces and mass sources. To demonstrate this, we proceed as
follows: The rate of change of energy of the system can also be
expressed by considering the system interacting with its
environment, in which case the internal interactions between
species (interaction forces, mass interconversion and
inter-species energy transfers) play no role. This viewpoint
gives,

\begin{eqnarray}
\sum\limits_{\iota}\frac{\mathrm{d}}{\mathrm{d}t}
\int\limits_{\Omega_0}\rho_0^\iota \left ( e^\iota + \frac{1}{2}
\Vert\bV+\bV^\iota\Vert^2 \right ) \mathrm{d}V =
\sum\limits_{\iota}\int\limits_{\Omega_0} \left(\rho_0^\iota\bg
\cdot\left(\bV+\bV^\iota\right) + \rho_0^\iota r^\iota
\right)\mathrm{d}V &
&\nonumber\\
+\sum\limits_{\iota}\int\limits_{\partial
\Omega_0}\left(\left(\bV+\bV^\iota\right)\cdot\bP^\iota -
\bM^\iota\left(e^\iota +\frac{1}{2}
\Vert\bV+\bV^\iota\Vert^2\right) -
\bQ^\iota\right)\cdot\bN\mathrm{d}A.\quad& & \label{energysys}
\end{eqnarray}

Since (\ref{energysum}) and (\ref{energysys}) are equivalent, it
follows that,

\begin{eqnarray}
\sum\limits_{\iota}\left(\int\limits_{\Omega_0}\left(\rho_0^\iota\bq^\iota\cdot(\bV+\bV^\iota)
+ \Pi^\iota\left(e^\iota +
\frac{1}{2}\Vert\bV+\bV^\iota\Vert^2\right)+\rho^\iota_0\tilde{e}^\iota\right)\mathrm{d}V
\right)= 0,
\end{eqnarray}

\noindent and on localizing this result,

\begin{eqnarray}
\sum\limits_{\iota}\left(\rho_0^\iota\bq^\iota\cdot(\bV+\bV^\iota)
+ \Pi^\iota\left(e^\iota +
\frac{1}{2}\Vert\bV+\bV^\iota\Vert^2\right)+\rho^\iota_0\tilde{e}^\iota\right)=
0. \label{energycond1}
\end{eqnarray}

\noindent This result relating the interaction energies to
interaction forces between species, their sources and relative
velocities, is identical to that obtained from classical mixture
theory \citep{TruesdellNoll:65}. Together with
(\ref{sourcebalance}) and (\ref{interforcebalance}) it
demonstrates that the present formulation is consistent with
mixture theory.

Equation (\ref{energyspecies1}) for the rate of change of energy
of a single species can be further simplified by applying the
Divergence Theorem and product rule, giving first,

\begin{eqnarray}
\int\limits_{\Omega_0}\left(\frac{\partial\rho_0^\iota}{\partial
t} \left(e^\iota + \frac{1}{2}\Vert\bV+\bV^\iota\Vert^2 \right) +
\rho^\iota_0\frac{\partial}{\partial t}\left(e^\iota +
\frac{1}{2}\Vert\bV+\bV^\iota\Vert^2 \right)\right) \mathrm{d}V
=\qquad
& &\nonumber\\
 \int\limits_{\Omega_0} \left(\rho_0^\iota\bg
\cdot\left(\bV+\bV^\iota\right) + \rho_0^\iota r^\iota +
\Pi^\iota\left(e^\iota
+ \frac{1}{2}\Vert\bV+\bV^\iota\Vert^2\right)+\rho^\iota_0\tilde{e}^\iota\right)\mathrm{d}V& & \nonumber \\
+\int\limits_{\Omega_0}\rho_0^\iota\bq^\iota\cdot(\bV+\bV^\iota)
\mathrm{d}V&
&\nonumber\\
+\int\limits_{\Omega_0}\left(\left(\bV+\bV^\iota\right)\cdot\Bnabla\cdot\bP^\iota
+
\bP^\iota\colon\Bnabla\left(\bV+\bV^\iota\right)\right)\mathrm{d}V& &\nonumber\\
-
\int\limits_{\Omega_0}\left(\Bnabla\cdot\left(\bM^\iota\right)\left(e^\iota
+\frac{1}{2} \Vert\bV+\bV^\iota\Vert^2\right)\right)\mathrm{d}V & &\nonumber\\
-\int\limits_{\Omega_0}\left( \left(\Bnabla e^\iota+
\left(\bV+\bV^\iota\right)\cdot\Bnabla\left(\bV+\bV^\iota\right)\right)\cdot\left(\bM^\iota\right)
- \Bnabla\cdot\bQ^\iota\right)\mathrm{d}V.
\end{eqnarray}

\noindent Using the balance of mass (\ref{massballocI}), balance
of linear momentum (\ref{ballinmomrefI}), and localizing the
result, we have,

\begin{eqnarray}
\rho^\iota_0\frac{\partial e^\iota}{\partial t} &=&
\bP^\iota\colon\Bnabla\left(\bV+\bV^\iota\right)-
\Bnabla\cdot\bQ^\iota + \rho_0^\iota r^\iota
+\rho^\iota_0\tilde{e}^\iota - \Bnabla e^\iota\cdot\bM^\iota
\label{energybalI}
\end{eqnarray}

\noindent Summing over $\iota$ gives,

\begin{eqnarray}
& &\sum\limits_{\iota}\rho^\iota_0\frac{\partial e^\iota}{\partial
t} =\nonumber\\
& &\qquad\sum\limits_{\iota}\left(
\bP^\iota\colon\dot{\bF}+\bP^\iota\colon\Bnabla\bV^\iota -
\Bnabla\cdot\bQ^\iota +\rho_0^\iota r^\iota +
\rho^\iota_0\tilde{e}^\iota - \Bnabla e^\iota\cdot\bM^\iota\right)
\label{energybaltot}
\end{eqnarray}

\noindent Substituting for $\sum\limits_\iota
\rho^\iota_0\tilde{e}^\iota$ from (\ref{energycond1}),

\begin{eqnarray}
\sum\limits_{\iota}\rho^\iota_0\frac{\partial e^\iota}{\partial t}
&=& \sum\limits_{\iota}\left(
\bP^\iota\colon\dot{\bF}+\bP^\iota\colon\Bnabla\bV^\iota -
\Bnabla\cdot\bQ^\iota +\rho_0^\iota r^\iota- \Bnabla
e^\iota\cdot\bM^\iota\right)\nonumber\\
&&-
\sum\limits_{\iota}\left(\rho_0^\iota\bq^\iota\cdot(\bV+\bV^\iota)
- \Pi^\iota\left(e^\iota +
\frac{1}{2}\Vert\bV+\bV^\iota\Vert^2\right)\right).
\label{energybaltot1}
\end{eqnarray}

This form of the balance of energy is most convenient for
combining with the entropy inequality leading to the
Clausius-Duhem form of the dissipation inequality.

\subsection{The entropy inequality: Clausius-Duhem form} \label{sect4.2}

Let $\eta^\iota$ be the entropy per unit mass of species $\iota$,
and $\theta$ the absolute temperature. The entropy production
inequality holds for the system as a whole. Accordingly, we write

\begin{eqnarray}
\sum\limits_{\iota}\frac{\mathrm{d}}{\mathrm{d}t}
\int\limits_{\Omega_0} \rho_0^\iota \eta^\iota \mathrm{d}V &\geq&
\sum\limits_{\iota}\int\limits_{\Omega_0}\left(
\Pi^\iota \eta^\iota + \frac{\rho_0^\iota r^\iota}{\theta}\right) \mathrm{d}V\nonumber\\
& & - \sum\limits_{\iota}\int\limits_{\partial \Omega_0}
\left(\bM^\iota \cdot \bN\eta^\iota + \frac{\bQ^\iota}{\theta}
\cdot \bN \right)\mathrm{d}A.
\end{eqnarray}
\noindent Applying the Divergence Theorem, using the mass balance
equation (\ref{massballocI}), and localizing the result, we have
the entropy inequality,

\begin{equation}
\sum\limits_{\iota}\rho_0^\iota\frac{\partial\eta^\iota}{\partial
t} \geq \sum\limits_{\iota}\left(\frac{\rho_0^\iota
r^\iota}{\theta} -\Bnabla\eta^\iota\cdot\bM^\iota -
\frac{\Bnabla\cdot\bQ^\iota}{\theta} +
\frac{\Bnabla\theta\cdot\bQ^\iota}{\theta^2}\right).
\label{entropyineq}
\end{equation}

Now, multiplying Equation (\ref{entropyineq}) by $\theta$,
subtracting it from Equation (\ref{energybaltot1}) and using
(\ref{ballinmomrefI}) for $\rho_0^\iota\bq^\iota$ gives,

\begin{eqnarray}
& &\sum\limits_{\iota}\rho^\iota_0\left(\frac{\partial
e^\iota}{\partial t} -\theta\frac{\partial\eta^\iota}{\partial
t}\right) +\sum\limits_{\iota}\Pi^\iota\left(e^\iota +
\frac{1}{2}\Vert\bV+\bV^\iota\Vert^2\right) +
\frac{\Bnabla\theta\cdot\bQ^\iota}{\theta}
\nonumber\\
& &+\sum\limits_{\iota} \left(\rho^\iota_0\frac{\partial}{\partial
t}\left(\bV+\bV^\iota\right) - \rho_0^\iota\bg -
\Bnabla\cdot\bP^\iota +
\Bnabla\left(\bV+\bV^\iota\right)\bM^\iota\right)\cdot\left(\bV+\bV^\iota\right)
\nonumber\\
& &-\sum\limits_{\iota}\left(\bP^\iota\colon\dot{\bF} -
\bP^\iota\colon\Bnabla\bV^\iota + \left(\Bnabla e^\iota -
\theta\Bnabla\eta^\iota\right) \cdot\bM^\iota\right)\leq 0
\label{redentropyineqfin}
\end{eqnarray}

\noindent Equation (\ref{redentropyineqfin}) is the reduced
entropy inequality---also referred to as the Clausius-Duhem
inequality---for growth processes.

\section{The kinematics of growth}\label{sect3bis}

The formulation up to this point has introduced some elements of
coupling between mass transport, mechanics and thermodynamics.
Mass transport and mechanics are further coupled due to the
kinematics of growth. Local volumetric changes take place as
species concentrations evolve. As concentration increases, the
material of a species swells, and conversely, shrinks as
concentration decreases. This observation has led to an active
field of study within the literature on biological growth
\citep{Skalak:81,SkalakHoger:96,TaberHumphrey:2001,LubardaHoger:02,AmbrosiMollica:2002}.
Our treatment follows in the same vein.

\subsection{The elasto-growth decomposition}\label{sect3bis.1}

Finite strain kinematics treats the total deformation gradient as
arising from a geometrically-necessary elastic deformation
accompanying growth, as well as a separate elastic deformation due
to an external stress. The deformation gradient is subject to a
split reminiscent of the classical decomposition of multiplicative
plasticity \citep{Bilbyetal:1957,Lee:1969}

\begin{figure}[ht]
\psfrag{A}{\small $\Omega_0$} \psfrag{B}{\small $\Omega^\ast$}
\psfrag{C}{\small $\Omega_t$} \psfrag{D}{\small $\Bvarphi$}
\psfrag{G}{\small $\bu^\ast$} \psfrag{E}{\small $\Bkappa$}
\psfrag{M}{\small $\bX$} \psfrag{I}{\small
$\bF^{\mathrm{g}^\iota}$} \psfrag{H}{\small
$\tilde{\bF}^{\mathrm{e}^\iota}$} \psfrag{J}{\small $\tilde{\bF}$}
\psfrag{Y}{\small $\bX^\ast$} \psfrag{K}{\small
$\bar{\bF}^\mathrm{e}$} \psfrag{X}{\small $\bx$} \psfrag{L}{\small
$\bF$} \centering
{\includegraphics[width=15cm]{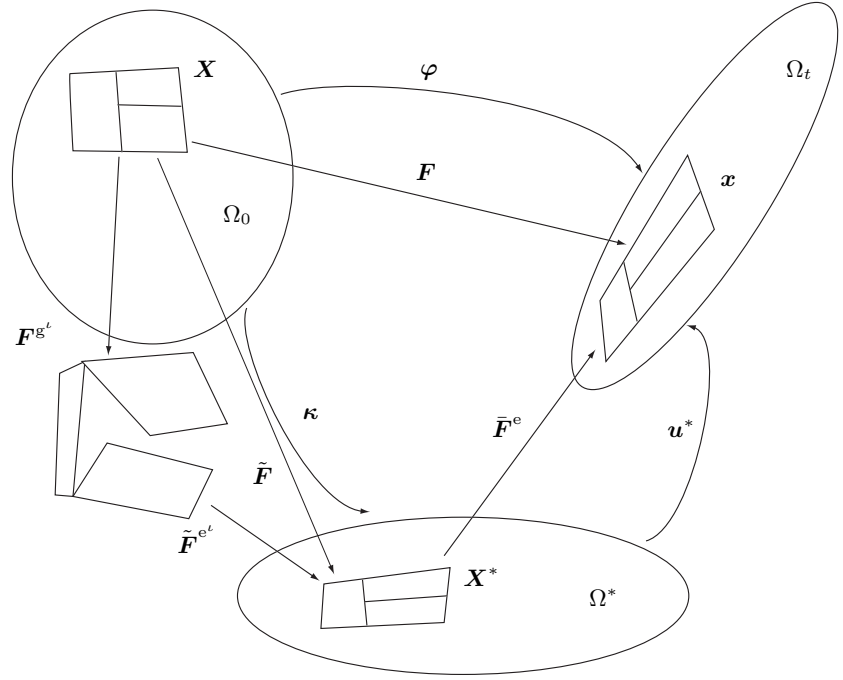}} \caption{The
kinematics of growth.} \label{growthkinematicsfig}
\end{figure}

At a continuum point the reference concentration of each species
admits the notion of an ``original'' state in which the
concentration of a species is $\rho_\mathrm{org}^\iota(\bX)$. This
is a state that may never be attained in a physical system.
However, if attained, the corresponding species would be
stress-free in the absence of deformation. Neglecting other
possible kinematics (such as plasticity) and microstructural
details, the set of quantities
$\{\rho_0^\mathrm{s},\dots,\rho_0^\omega\}$, and the temperature,
$\theta$, fully specify the reference state of the material at a
point. As mass transport alters the reference density to its value
$\rho_0^\iota(\bX,t)$, the species swells if $\rho_0^\iota >
\rho_\mathrm{org}^\iota$, and shrinks if $\rho_0^\iota <
\rho_\mathrm{org}^\iota$. Assuming that these volume changes are
isotropic leads to the following growth kinematics: For each
species, one can define a ``growth deformation gradient tensor'',
$\bF^{\mathrm{g}^\iota} :=
\frac{\rho_0^\iota}{\rho_\mathrm{org}^\iota}{\bf 1}$, where ${\bf
1}$ is the second-order isotropic tensor. The tensor
$\bF^{\mathrm{g}^\iota}$ is analogous to the plastic deformation
gradient of multiplicative plasticity. As the ratio
$\rho_0^\iota/\rho_\mathrm{org}^\iota$ is a local quantity,
$\bF^{\mathrm{g}^\iota}$ varies pointwise and adjacent
neighborhoods will, in general, be incompatible due to the action
of $\bF^{\mathrm{g}^\iota}$ alone. However, further elastic
deformation, $\tilde{\bF}^{\mathrm{e}^\iota}$ occurs to ensure
compatibility, leading to an internal stress, that can in general
be different for each species. The action of these kinematic
tangent maps can be conceived of in the absence of external
stress. With an external stress, there is further elastic
deformation, $\bar{\bF}^\mathrm{e}$, common to all species. This
sequence of maps is pictured in Figure \ref{growthkinematicsfig}.
The kinematic relations are:
\begin{equation}
\bF =
\bar{\bF}^\mathrm{e}\tilde{\bF}^{\mathrm{e}^\iota}\bF^{\mathrm{g}^\iota},\quad
\bF^{\mathrm{g}^\iota} =
\frac{\rho_0^\iota}{\rho_\mathrm{org}^\iota}{\bf 1}.
\label{growthkinematicseq}
\end{equation}

\noindent Clearly, the elastic deformation gradients can be
combined to write $\bF^{\mathrm{e}^\iota} =
\bar{\bF}^\mathrm{e}\tilde{\bF}^{\mathrm{e}^\iota}$, the ``total''
elastic deformation gradient of species $\iota$.

\section{Restrictions on constitutive relations from the
Clausius-Duhem inequality} \label{sect5}

As is the practice in field theories of continuum physics, we use
the Clausius-Duhem inequality (\ref{redentropyineqfin}) to obtain
restrictions on the constitutive relations. We begin with very
general assumptions on the specific internal energy of each
species: $ e^\iota = \hat{e}^\iota(\bF^{\mathrm{e}^\iota},
\eta^\iota, \rho_0^\iota)$. Then, applying the chain rule and
regrouping some terms, (\ref{redentropyineqfin}) becomes
\begin{eqnarray}
& &\sum\limits_{\iota}\left(\frac{\partial
e^\iota}{\partial\bF^{\mathrm{e}^\iota}}-\bP^\iota\bF^{\mathrm{g}^{\iota\mathrm{T}}}\right)\colon\dot{\bF}^{\mathrm{e}^\iota}
+ \sum\limits_{\iota}\left(\frac{\partial e^\iota}{\partial
\eta^\iota}-\theta\right)\frac{\partial\eta^\iota}{\partial t}\nonumber\\
&+&\sum\limits_{\iota} \left(\rho^\iota_0\frac{\partial}{\partial
t}\left(\bV+\bV^\iota\right) - \rho_0^\iota\bg -
\Bnabla\cdot\bP^\iota +
\Bnabla\left(\bV+\bV^\iota\right)\bM^\iota\right)\cdot(\bV^\iota+\bV)\nonumber\\
&+&\sum\limits_{\iota}\left(\rho^\iota_0\bF^{-\mathrm{T}}\left(\Bnabla
e^\iota -
\theta\Bnabla\eta^\iota\right)\right)\cdot\bV^\iota\nonumber\\
&+&\sum\limits_{\iota}\Pi^\iota\left(e^\iota +
\frac{1}{2}\Vert\bV+\bV^\iota\Vert^2\right) +
\frac{\Bnabla\theta\cdot\bQ^\iota}{\theta}\nonumber\\
&+& \sum\limits_{\iota}\rho^\iota_0\frac{\partial
e^\iota}{\partial\rho^\iota_0}\frac{\partial
\rho^\iota_0}{\partial t} -
\sum\limits_{\iota}\bP^\iota\colon(\Bnabla\bV^\iota +
\bF^{\mathrm{e}^\iota}\dot{\bF}^{\mathrm{g}^\iota}) \leq 0.
\label{redentropyineq1}
\end{eqnarray}

\noindent Inequality (\ref{redentropyineq1}) represents a
fundamental restriction upon the physical processes during
biological growth. Any constitutive relations that are prescribed
must satisfy this restriction, as is well-known
\citep{TruesdellNoll:65}. Guided by (\ref{redentropyineq1}), we
prescribe the following constitutive relations in classical form:
\begin{eqnarray}
&&\bP^\iota\bF^{\mathrm{g}^\iota\mathrm{T}} = \rho_0^\iota
\frac{\partial e^\iota}{\partial\bF^{\mathrm{e}^\iota}}
\label{stress-constrelI}\\
\nonumber\\
&&\theta =  \frac{\partial e^\iota}{\partial \eta^\iota},\;\forall\,\iota\label{temp-constrelI}\\
\nonumber\\
&&\rho^\iota_0\bV^\iota =\nonumber\\
& &
-\frac{\tilde{\bD}^\iota}{\rho^\iota_0}\left(\rho^\iota_0\frac{\partial\bV}{\partial
t} - \rho_0^\iota\bg - \Bnabla\cdot\bP^\iota +
\left(\Bnabla\bV\right)\bM^\iota+\rho^\iota_0\bF^{-\mathrm{T}}\left(\Bnabla
e^\iota -
\theta\Bnabla\eta^\iota\right)\right),\label{VIconstrel1}\\
&&\mathrm{where}\;\bw\cdot\tilde{\bD}^\iota\bw
\ge 0,\;\;\forall\, \bw \in\mathbb{R}^3\nonumber\\
\nonumber\\
&&\bQ^\iota = -{\bK}^\iota\Bnabla\theta,\;\bw\cdot{\bK}^\iota\bw
\ge 0,\;\;\forall\, \bw \in\mathbb{R}^3\label{qconstrelI}
\end{eqnarray}

\noindent With the constitutive relations
(\ref{stress-constrelI}--\ref{qconstrelI}) ensuring the
non-positiveness of certain terms the entropy inequality is
further reduced to
\begin{eqnarray}
& & \sum\limits_{\iota}\left(\rho^\iota_0\frac{\partial
e^\iota}{\partial\rho^\iota_0}\frac{\partial
\rho^\iota_0}{\partial t} - \bP^\iota\colon(\Bnabla\bV^\iota +
\bF^{\mathrm{e}^\iota}\dot{\bF}^{\mathrm{g}^\iota})\right)\nonumber\\
&&+\sum\limits_{\iota}\left(
\rho^\iota_0\bV^\iota\cdot\left(\frac{\partial\bV^\iota}{\partial
t} +
\left(\Bnabla\bV^\iota\right)\bF^{-1}\bV^\iota\right)+\Pi^\iota\left(e^\iota
+ \frac{1}{2}\Vert\bV+\bV^\iota\Vert^2\right)\right)\nonumber\\
&&+\sum\limits_{\iota} \left(\rho^\iota_0\frac{\partial}{\partial
t}\left(\bV+\bV^\iota\right) - \rho_0^\iota\bg -
\Bnabla\cdot\bP^\iota +
\Bnabla\left(\bV+\bV^\iota\right)\bM^\iota\right)\cdot\bV \le 0.
\label{dissipation1}
\end{eqnarray}

\noindent The left hand-side of (\ref{dissipation1}) is the
dissipation, $\sD$, a quantity that we will return to below.

Equation (\ref{stress-constrelI}) specifies a constitutive
relation for $\bP^\iota\bF^{\mathrm{g}^{\iota\mathrm{T}}}$, which
is a truly elastic stress. Equation (\ref{temp-constrelI}) implies
a uniform temperature in each species, and corresponds with the
definition usually employed for temperature in thermal physics.
The heat flux in species $\iota$ is given by the product of a
positive semi-definite conductivity tensor, ${\bK}^\iota$ and the
temperature gradient, which is the Fourier Law of heat conduction.
Equation (\ref{VIconstrel1}) requires more detailed discussion
which appears below.

\subsection{Constitutive relations for fluxes}\label{sect5.1}

Turning to (\ref{VIconstrel1}), we first point out that since
$\bM^\iota = \rho_0^\iota\bF^{-1}\bV^\iota$, this is an implicit
relation for $\bV^\iota$. Rewriting it as an explicit one for
$\rho^\iota_0\bV^\iota$ we have,
\begin{eqnarray}
\rho^\iota_0\bV^\iota =& &\left(\bone +
\frac{\tilde{\bD}^\iota\Bnabla\bV\bF^{-1}}{\rho^\iota_0}\right)^{-1}\frac{\tilde{\bD}^\iota}{\rho^\iota_0}\nonumber\\
& &\cdot\left\{-\left(\rho^\iota_0\frac{\partial\bV}{\partial t} -
\rho_0^\iota\bg - \Bnabla\cdot\bP^\iota +
\rho^\iota_0\bF^{-\mathrm{T}}\left(\Bnabla e^\iota -
\theta\Bnabla\eta^\iota\right)\right)\right\}. \label{VIconstrel2}
\end{eqnarray}

\noindent The constitutive relation for flux, $\bM^\iota =
\rho^\iota_0\bF^{-1}\bV^\iota$ is then obtained:
\begin{eqnarray}
\bM^\iota & &=\underbrace{\bF^{-1}\left(\bone +
\frac{\tilde{\bD}^\iota\Bnabla\bV\bF^{-1}}{\rho^\iota_0}\right)^{-1}\frac{\tilde{\bD}^\iota}{\rho^\iota_0}\bF^{-\mathrm{T}}}_{\bD^\iota}\nonumber\\
&\cdot&\left\{\underbrace{-\left(\rho^\iota_0\bF^\mathrm{T}\frac{\partial\bV}{\partial
t} - \rho_0^\iota\bF^\mathrm{T}\bg -
\bF^\mathrm{T}\Bnabla\cdot\bP^\iota + \rho^\iota_0\left(\Bnabla
e^\iota -
\theta\Bnabla\eta^\iota\right)\right)}_{\boldmath{\sF}^\iota}\right\}.
\label{MIconstrel}
\end{eqnarray}

As is common in descriptions of mass transport, the tensor
delineated as $\bD^\iota$ will be referred to as the mobility
tensor of species $\iota$. Recall that in (\ref{VIconstrel1}) we
have taken $\tilde{\bD}^\iota$ to be positive
semi-definite\footnote{If
$\Vert\tilde{\bD}^\iota\Bnabla\bV\bF^{-1}/\rho^\iota_0\Vert << 1$,
we have $\bD^\iota \approx \tilde{\bD}^\iota$.}. The flux is thus
written as the product of a mobility tensor and a thermodynamic
driving force, $\boldmath{\sF}^\iota$. This is the Nernst-Einstein
relation. We proceed now to examine the four separate terms in the
thermodynamic driving force:
\begin{equation}
\boldmath{\sF}^\iota =
-\rho^\iota_0\bF^\mathrm{T}\frac{\partial\bV}{\partial t} +
\rho_0^\iota\bF^\mathrm{T}\bg +
\bF^\mathrm{T}\Bnabla\cdot\bP^\iota - \rho^\iota_0\left(\Bnabla
e^\iota - \theta\Bnabla\eta^\iota\right) \label{drivingforceI}
\end{equation}

\noindent The first two terms respectively represent the
influences of inertia and body force. Thus, the inertial effect is
to drive species $\iota$ in the opposite direction to the body's
acceleration. The body force's influence is directed along itself.
The third term represents the stress divergence effect. In the
case of a non-uniform partial stress, $\bP^\iota$, there exists a
thermodynamic driving force for transport along $\bP^\iota$. We
demonstrate this effect for the case of the fluid species in
Section \ref{sect5.2} below, for which it translates to the more
intuitive notion of transport along a fluid pressure gradient.

The fourth term in $\boldmath{\sF}^\iota$ admits the following
interpretation: The Legendre transformation $\psi^\iota = e^\iota
- \theta\eta^\iota$ allows one to rewrite $\Bnabla e^\iota -
\theta\Bnabla\eta^\iota$ as $\Bnabla\psi^\iota\vert_\theta$ (at
uniform temperature), where $\psi^\iota$ is the mass-specific
Helmholtz free energy. An assumption inherent in the development
that began in Section \ref{sect2} is that any mass entering or
leaving $\Omega_0$ at a point $\bX$ on the boundary,
$\partial\Omega_0$, has the field values
$\rho_0^\iota,e^\iota,\eta^\iota,\theta$, and $\psi^\iota$
corresponding to $\bX$. Likewise, the incremental mass of species
$\iota$ created or absorbed via the source/sink $\Pi^\iota$ at
$\bX$ has the field values of that point. Consider a sufficiently
small neighborhood of a point, say
$\mathsf{N}(\bX)\subset\Omega_0$. Changing the mass of species
$\iota$ in $\mathsf{N}(\bX)$ by $\delta\mathsf{m}^\iota$ units
causes a change in the Helmholtz free energy of $\iota$ in
$\mathsf{N}(\bX)$ by $\delta\Psi^\iota =
\psi^\iota\delta\mathsf{m}^\iota$. By definition therefore,
$\psi^\iota =
\partial\Psi^\iota/\partial\mathsf{m}^\iota$. This derivative gives the \emph{chemical
potential}, $\mu^\iota$, of the transported species, $\iota$.
Thus, we have $\mu^\iota = e^\iota - \theta\eta^\iota$, and
$\Bnabla e^\iota -\theta\Bnabla\eta^\iota =
\Bnabla\mu^\iota\vert_\theta$. This last term in
${\boldmath{\sF}}^\iota$ thus represents the thermodynamic driving
force due to a chemical potential gradient.

It has recently come to our attention that the constitutive
relation for flux (\ref{MIconstrel}) is precisely the result
arrived at by \citet{DeGrootMazur:1984}, including the
identification of the chemical potential gradient term. However,
their approach involves a slightly different application of the
Second Law, and a less detailed treatment of the mechanics.

The gradient of internal energy in (\ref{drivingforceI}) leads to
a strain gradient-dependent term. A concentration gradient-driven
term arises from the gradient of mixing entropy. Together with the
other terms that were remarked upon above, they represent a
complete thermodynamic formulation of coupled mass transport and
mechanics. This is the central result of our paper.

\subsection{Transport of the fluid species: The example of an ideal fluid}\label{sect5.2}

Consider the stress divergence term
$\bF^\mathrm{T}\Bnabla\cdot\bP^\iota$. An elementary calculation
gives
\begin{equation}
\bF^\mathrm{T}\Bnabla\cdot\bP^\iota =
\Bnabla\cdot\left(\bF^\mathrm{T}\bP^\iota\right) -
\Bnabla\bF^\mathrm{T}\colon\bP^\iota. \label{stressdivI}
\end{equation}

\noindent In indicial form, where lower/upper case indices are for
components of quantities in the current/reference configuration
respectively, this relation is
\begin{displaymath}
F_{iK}P^\iota_{iJ,J} = \left(F_{iK}P^\iota_{iJ}\right)_{,J} -
F_{iK,J}P^\iota_{iJ}.
\end{displaymath}

\noindent For an ideal fluid, supporting only an isotropic Cauchy
stress, $p\bone$, we have $\bP^\mathrm{f} =
\mathrm{det}(\bF)p\bF^{-\mathrm{T}}$, where $p$ is positive in
tension. The arguments that follow assume this case. (The more
general case of a non-ideal, viscous fluid will merely have
additional terms from the viscous Cauchy stress.) The stress
divergence term is
\begin{equation}
\bF^\mathrm{T}\Bnabla\cdot\bP^\mathrm{f} =
\Bnabla\left(\mathrm{det}(\bF)p\right) -
\Bnabla\bF^\mathrm{T}\colon\bF^{-\mathrm{T}}\mathrm{det}(\bF)p,
\end{equation}

\noindent demonstrating the appearance of a hydrostatic
stress-driven contribution to ${\boldmath{\sF}}^\mathrm{f}$. This
is Darcy's Law for transport of a fluid down a pressure gradient.

For the special case of a compressible, ideal fluid we have
$e^\mathrm{f} =
\bar{e}^\mathrm{f}(\eta^\mathrm{f},\bar{\rho}^\mathrm{f})$; i.e.,
the fluid stores strain energy as a function of its \emph{current,
intrinsic} density. Fluid saturation conditions hold in biological
tissue, for which case the fluid volume fraction, $f^\mathrm{f}$,
is simply the pore volume fraction. Recall from Section
\ref{sect2} that the individual species deform with the common
deformation gradient $\bF$. Therefore the pores deform
\emph{homogeneously} with the surrounding solid phase. Physically
this corresponds to the pore size being smaller than the scale at
which the homogenization assumption of a continuum theory holds.
Momentarily ignoring changes in reference concentration of the
fluid, we have $\bF^{\mathrm{e}^\mathrm{f}} = \bF$.  Then, since
$\rho^\mathrm{f}_0 = \bar{\rho}^\mathrm{f}_0 f^\mathrm{f}$, we can
write $\hat{e}^\mathrm{f}(\bF,\eta^\mathrm{f},\rho^\mathrm{f}_0) =
\hat{e}^\mathrm{f}(\bF,\eta^\mathrm{f},\bar{\rho}^\mathrm{f}_0
f^\mathrm{f}) =
\bar{e}^\mathrm{f}(\eta^\mathrm{f},\bar{\rho}^\mathrm{f}_0/\mathrm{det}\bF)=
\bar{e}^\mathrm{f}(\eta^\mathrm{f},\bar{\rho}^\mathrm{f})$. In
this case a simple calculation shows that the hydrostatic pressure
is
\begin{displaymath}
p =
-\frac{\bar{\rho}^\mathrm{f}}{\mathrm{det}(\bF)}\frac{\partial\bar{e}^\mathrm{f}}{\partial\bar{\rho}^\mathrm{f}},
\end{displaymath}

\noindent and the stress divergence term is
\begin{displaymath}
\bF^\mathrm{T}\Bnabla\cdot\bP^\mathrm{f} =
-\Bnabla\left(\bar{\rho}^\mathrm{f}\frac{\partial\bar{e}^\mathrm{f}}{\partial\bar{\rho}^\mathrm{f}}\right)
+
\Bnabla\bF^\mathrm{T}\colon\bF^{-\mathrm{T}}\bar{\rho}^\mathrm{f}\frac{\partial\bar{e}^\mathrm{f}}{\partial\bar{\rho}^\mathrm{f}}.
\end{displaymath}

\subsection{The Eshelby stress as a thermodynamic driving
force}\label{sect5.3}

Combining the stress divergence and chemical potential gradient
contributions to the driving force for any species, and using the
mass-specific Helmholtz free energy, $\psi^\iota$, we write,

\begin{equation}
\bF^\mathrm{T}\Bnabla\cdot\bP^\iota - \rho^\iota_0\left(\Bnabla
e^\iota - \theta\Bnabla\eta^\iota\right) =
\Bnabla\cdot\left(\bF^\mathrm{T}\bP^\iota\right) -
\Bnabla\bF^\mathrm{T}\colon\bP^\iota
 - \rho^\iota_0\Bnabla\psi^\iota\vert_\theta.
\end{equation}

\noindent Regrouping terms this expression is
\begin{equation}
-\Bnabla\cdot\underbrace{\left(\rho^\iota_0\psi^\iota\vert_\theta\bone
- \bF^\mathrm{T}\bP^\iota\right)}_{\mbox{Eshelby
stress},\;\BXi^\iota} +
\left(\Bnabla\rho^\iota_0\right)\psi^\iota_0\vert_\theta -
\Bnabla\bF^\mathrm{T}\colon\bP^\iota.
\end{equation}

Thus, the divergence of the well-known Eshelby stress tensor is
also among the driving forces for mass transport. Also observe the
presence of a strain gradient-dependent driving force,
$-\Bnabla\bF^\mathrm{T}\colon\bP^\iota$ in the developments of
Sections \ref{sect5.2} and \ref{sect5.3}, independent of the
pressure gradient term for the fluid species.

\noindent{\bf Remark 1}: The final version of the dissipation
inequality (\ref{dissipation1}), and the mass balance equation can
be manipulated to restrict the mathematical form of the mass
source. It is common to make the mass source depend upon the
strain energy density \citep{HarriganHamilton:1993} while
respecting the restriction imposed by the dissipation inequality.
This form is often used while modelling hard tissue. Such an
approach leads to strain-mediated mass transport. However, with a
strain-independent source, strain-mediated (or stress-mediated)
mass transport would not be obtained with such a formulation.

\noindent{\bf Remark 2}: We expect that evaluation of the
dissipation, $\sD$, using (\ref{dissipation1}) from field
quantities in a boundary value problem will provide a test of
soundness, and if necessary indications for improvement, of our
constitutive models.

\noindent{\bf Remark 3}: Since soft biological tissues usually
demonstrate rate-dependent response, it has been common to employ
a solid viscoelastic constitutive model for them. This approach
fits within our framework, with a modification of the internal
energy to include its dependence upon internal variables that
represent the viscoelastic stress-like parameters. However, a more
physiologically-valid model may be one with a purely hyperelastic
solid phase, and a viscous fluid. In such a composite model the
rate-dependent behavior would arise from the fluid.

\noindent{\bf Remark 4}: The constitutive relations
(\ref{stress-constrelI}) and (\ref{MIconstrel}) respectively
specify the partial stress, $\bP^\iota$, and flux, $\bM^\iota$, of
a species. The flux also implies the relative velocity,
$\bV^\iota$. The velocity of the solid phase, $\bV$ is obtained
from the local form of the balance of linear momentum for the
system (\ref{linmombalsys}). With all these quantities known, the
individual interaction forces between species,
$\rho^\iota_0\bq^\iota$, can be obtained from
(\ref{ballinmomrefI}). They are, however, not needed while solving
for the balance of linear momentum of the system.

\section{A numerical example} \label{sect6}

The theory developed in Sections \ref{sect2}--\ref{sect5} has been
implemented in a computational formulation, retaining much of the
complexity of the coupled balance laws and constitutive relations.
For realistic soft tissue material parameters, the contribution of
the fluxes and interaction forces between species to the balance
of linear momentum of the composite tissue is negligible. This
simplification has been used. As a preliminary demonstration of
the theory\footnote{This numerical section has been included
mainly for completeness of this theoretical paper. A separate
paper, currently in preparation, will present the computational
formulation and contain a detailed examination of a number of
initial and boundary value problems for growth.}, we present a
computation of the coupled physics in the early stages of uniaxial
extension of a cylindrical soft tissue specimen. The motivation
for this model problem comes from our experimental model of
engineered, functional tendon constructs grown \emph{in vitro},
having the same cylindrical geometry. The experimental aspects of
our broad-based project on soft tissue growth are described
elsewhere \citep{Calveetal:2003}. In addition to engineering
scaffold-less tendon constructs from neonatal rat fibroblast
cells, we have the ability to impose a range of mechanical,
chemical, nutritional and electrical stimuli on them and study the
tissue's response. Besides modelling these experiments, the
mathematical formulation described here presents researchers with
a vehicle for testing scenarios and framing hypotheses that can be
experimentally-validated in our laboratory, thereby driving the
experimental studies.

\subsection{Material models and parameters}\label{sect6.1}

\begin{figure}[ht]
\psfrag{A}{$a$} \psfrag{B}{$b$} \psfrag{C}{$c$} \centering
{\includegraphics[width=3.5cm]{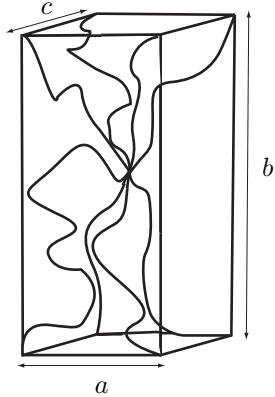}} \caption{Worm-like
chains grouped into an initially anisotropic eight-chain model.}
\label{eightchain}.
\end{figure}

The engineered tendon construct is $12$ mm in length and
$1\;\mathrm{mm}^2$ in area. In this paper an internal energy
density for the solid phase based upon the worm-like chain model
is used. The reader is directed to \citet{Riefetal:97} and
\cite{Bustamanteetal:2003} where the one-dimensional version of
this model has been applied to long chain molecules. It has been
described and implemented into an anisotropic representative
volume element by \citet{Bischoffetal:2002}, and is summarized
here. The internal energy density of a single constituent chain of
an eight-chain model (Figure \ref{eightchain}) is,
\begin{eqnarray}
\bar{\rho}_0^\mathrm{s}\hat{e}^\mathrm{s}(\bF^{\mathrm{e}^\mathrm{s}},\rho_0^\mathrm{s},\eta^\mathrm{s})
&=& \frac{N k \theta}{4 A}\left(\frac{r^2}{2L} +
\frac{L}{4(1-r/L)} -
\frac{r}{4}\right)\nonumber\\
&-&\frac{N k \theta}{4\sqrt{2L/A}}\left(\sqrt{\frac{2A}{L}} +
\frac{1}{4(1 - \sqrt{2A/L})} -\frac{1}{4} \right)\log(\lambda_1^{a^2}\lambda_2^{b^2}\lambda_3^{c^2})\nonumber\\
&+& \frac{\gamma}{\beta}({J^{\mathrm{e}^\mathrm{s}}}^{-2\beta} -1)
+ 2\gamma{\bf 1}\colon\bE^{\mathrm{e}^\mathrm{s}} \label{wlcmeq}
\end{eqnarray}
\noindent Here, $N$ is the density of chains, $k$ is the Boltzmann
constant, $r$ is the end-to-end length of a chain, $L$ is the
fully-extended length, and $A$ is the persistence length that
measures the degree to which the chain departs from a straight
line. The preferred orientation of tendon collagen is described by
an anisotropic unit cell with sides $a,b$ and $c$---see Figure
\ref{eightchain}. All lengths in this model have been rendered
non-dimensional (Table \ref{mattab}) by dividing by the link
length in a chain.

The elastic stretches along the unit cell axes are respectively
denoted $\lambda^\mathrm{e}_1,\lambda^\mathrm{e}_2$ and
$\lambda^\mathrm{e}_3$, and $\bE^{\mathrm{e}^\mathrm{s}} =
\frac{1}{2}(\bC^{\mathrm{e}^\mathrm{s}} - {\bf 1})$ is the elastic
Lagrange strain. The factors $\gamma$ and $\beta$ control bulk
compressibility. The end-to-end length is given by

\begin{equation}
r =
\frac{1}{2}\sqrt{a^2\lambda_1^{\mathrm{e}^2}+b^2\lambda_2^{\mathrm{e}^2}+c^2\lambda_3^{\mathrm{e}^2}},\quad
\lambda_I^{\mathrm{e}} = \sqrt{\bN_I\cdot\bC^{\mathrm{e}}\bN_I}
\label{rwlcm}
\end{equation}

Preliminary mechanical tests of the engineered tendon have been
carried out in our laboratory but, at this stage, the worm-like
chain model has not been calibrated to these tests. Instead,
published data for the worm-like chain, obtained by calibrating
against rat cardiac tissue  \citep{Bischoffetal:2002}, has been
employed.

The fluid phase was modelled as an ideal, nearly-incompressible
fluid:
\begin{equation}
\bar{\rho}^\mathrm{f}_0\hat{e}^\mathrm{f}(\bF^{\mathrm{e}^\mathrm{f}},\rho_0^\mathrm{f},\eta^\mathrm{f})
=
\frac{1}{2}\kappa(\mathrm{det}(\bF^{\mathrm{e}^\mathrm{f}})-1)^2,
\end{equation}

\noindent where $\kappa$ is the fluid bulk modulus.

Only a solid and a fluid phase were included for the tissue. Low
values were chosen for the mobilities of the fluid
\citep{Swartzetal:99} with respect to the solid phase (see Table
\ref{mattab}). In order to demonstrate growth, the solid phase
must have a source term, $\Pi^\mathrm{s}$ (Section \ref{sect2}),
and the only other phase, the fluid, must have $\Pi^\mathrm{f} =
-\Pi^\mathrm{s}$. Therefore, contrary to the case made in Section
\ref{sect2}, a non-zero value of the fluid source,
$\Pi^\mathrm{f}$, was assumed. A form motivated by first-order
reactions was used:
\begin{equation}
\Pi^\mathrm{f} = -k^\mathrm{f}(\rho_0^\mathrm{f} -
\rho_{0_\mathrm{ini}}^\mathrm{f}),\quad \Pi^\mathrm{s} =
-\Pi^\mathrm{f}, \label{piform}
\end{equation}

\noindent where $k^\mathrm{f}$ is the reaction rate, and
$\rho_{0_\mathrm{ini}}^\mathrm{f}$ is the initial fluid
concentration. This term acts as a source for the solid when
$\rho_0^\mathrm{f}
> \rho_{0_\mathrm{ini}}^\mathrm{f}$, and a sink when
$\rho_0^\mathrm{f} < \rho_{0_\mathrm{ini}}^\mathrm{f}$.

In a very simple approximation, the fluid's mixing entropy was
written as
\begin{equation}
\eta^\mathrm{f}_\mathrm{mix} =
-\frac{k}{\sM^\mathrm{f}}\log\frac{\rho_0^\mathrm{f}}{\rho_0}.
\label{mixentropy}
\end{equation}

\noindent Recall that in the notation of Section \ref{sect2},
$\sM^\mathrm{f}$ is the fluid's molecular weight.

\begin{table}[ht]
\caption{Material parameters used in the analysis} \label{mattab}
\begin{tabular}{lcll}
\hline
\multicolumn{1}{c}{Parameter} & Symbol & Value & Units\\
\hline
Chain density & $N$ & $7\times 10^{21}$ & $\mathrm{m}^{-3}$\\
Temperature& $\theta$  & $310.0$ & K\\
Persistence length & $A$ & $1.3775$ & --\\
Fully-stretched length & $L$ & $25.277$ & --\\
Unit cell axes & $a,\;b,\;,c$ & $9.2981,\;12.398,\;6.1968$ & --\\
Bulk compressibility factors & $\gamma,\;\beta$ & $1000,\; 4.5$ & --\\
Fluid bulk modulus &$\kappa$ & $1$ & GPa\\
Fluid mobility tensor components& $D_{11},\;D_{22},\;D_{33}$ & $1\times 10^{-8},\;1\times 10^{-8},\;1\times 10^{-8}$ &$\mathrm{m}^{-2}\mathrm{sec}$\\
Fluid conversion reaction rate & $k^\mathrm{f}$ & $-1.\times 10^{-7}$ & $\mathrm{sec}^{-1}$\\
Gravitational acceleration & $\bg$ & $9.81$ & $\mathrm{m}.\mathrm{sec}^{-2}$\\
Molecular weight of fluid &$\sM^\mathrm{f}$& $2.9885\times 10^{-23}$ & $\mathrm{kg}$\\
\hline
\end{tabular}
\end{table}

\subsection{Boundary and initial conditions; coupled solution method}\label{sect6.2}

Boundary conditions for mass transport consisted of the specified
fluid concentration at all external surfaces of the cylinder. This
value was fixed at $500\,\mathrm{kg.m}^{-3}$. With these boundary
conditions the fluid flux normal to surfaces of the specimen is
determined by solving the initial and boundary value problem. The
bottom planar surface was fixed in the $\be_3$ direction and a
displacement was applied at the top surface, also in the $\be_3$
direction, to give a nominal strain rate of
$0.05\,\mathrm{sec}^{-1}$ in the $\be_3$ direction. This is the
only mechanical load on the problem. Initial conditions were
$\rho_0^\mathrm{f}(\bX,0) =
500\,\mathrm{kg.m}^{-3},\;\rho_0^\mathrm{s}(\bX,0) =
500\,\mathrm{kg.m}^{-3}$, and for the mechanical problem,
$\bu(\bX,0) = \bzero,\,\bV(\bX,0) = \bzero$.

The coupled problem was solved by a staggered scheme based upon
operator splits \citep{Armero:99,Garikipatietal:01}. The details
will be presented in a future communication that will focus upon
computational aspects and numerical examples. Here we only mention
that the staggered scheme consists of identifying the displacement
and species concentrations as primitive variables associated with
the mechanical and mass transport problems. The mechanical problem
is solved holding the concentrations fixed. The resulting
displacement field is then held constant to solve the mass
transport problem. The transient solution is obtained for
mechanics using energy-momentum conserving schemes
\citep{SimoTarnow:1992b,SimoTarnow:1992a,Gonzalezphd:1996}, and
for mass transport using the Backward Euler Method. Hexahedral
elements are employed, combined with nonlinear projection methods
\citep{SimoTaylorPister:85} to treat the near-incompressibility
imposed by the fluid. The numerical formulation has been
implemented within the nonlinear finite element program, FEAP
\citep{FEAPmanual}.

\subsection{Evolution of stress and flux}\label{sect6.3}

The following contour plots represent the stress, and various
contributions to the total flux in the early stages of loading of
the model problem. Symmetry has been employed to model a single
quadrant of the cylinder.

\begin{figure}[ht]
\begin{minipage}[t]{7.5cm}
{\includegraphics[width=7.5cm]{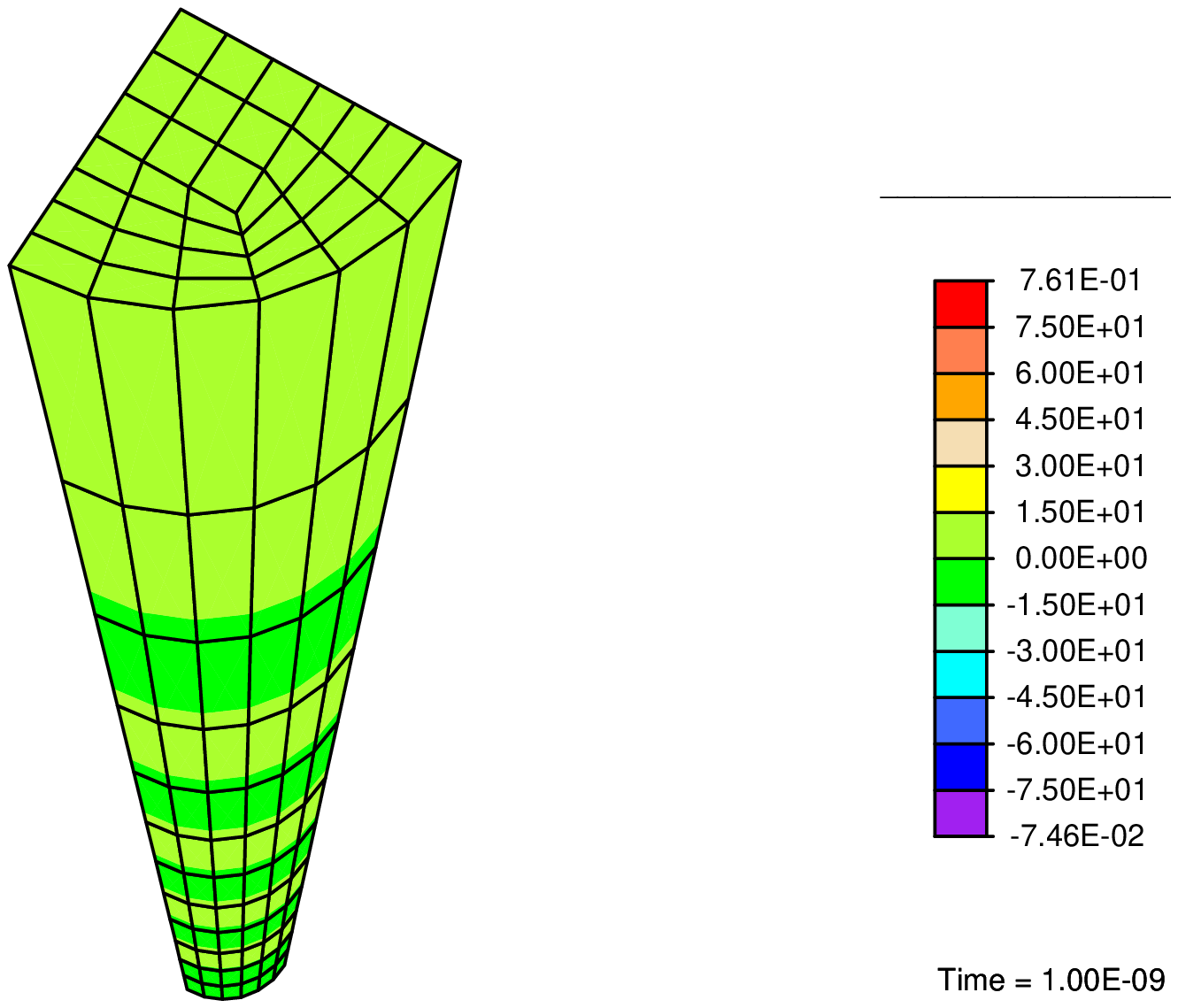}} \hskip 3cm (a)
\end{minipage}
\begin{minipage}[t]{7.5cm}
{\includegraphics[width=7.5cm]{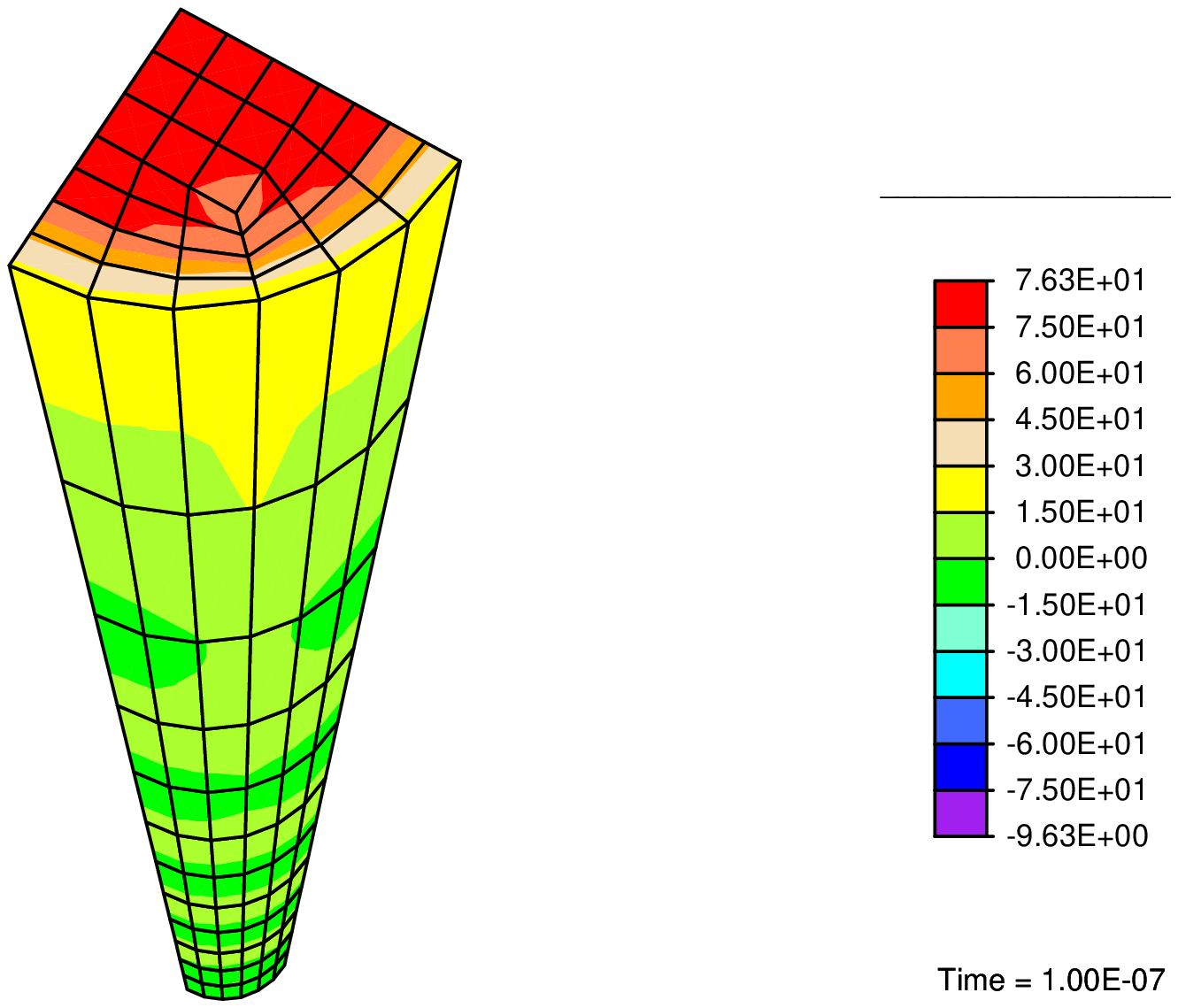}} \hskip 3cm (b)
\end{minipage}
\caption{Longitudinal Cauchy stress, $\sigma_{33}$ (Pa) at $1
\,\mathrm{nanosec.}$ and $100\,\mathrm{nanosec.}$ after the
beginning of loading.} \label{stressfig}
\end{figure}

The longitudinal stress, $\sigma_{33}$ in Figure \ref{stressfig}
arises from the stretch and the evolution in concentration.

\begin{figure}[ht]
\begin{minipage}[t]{7.5cm}
{\includegraphics[width=7.5cm]{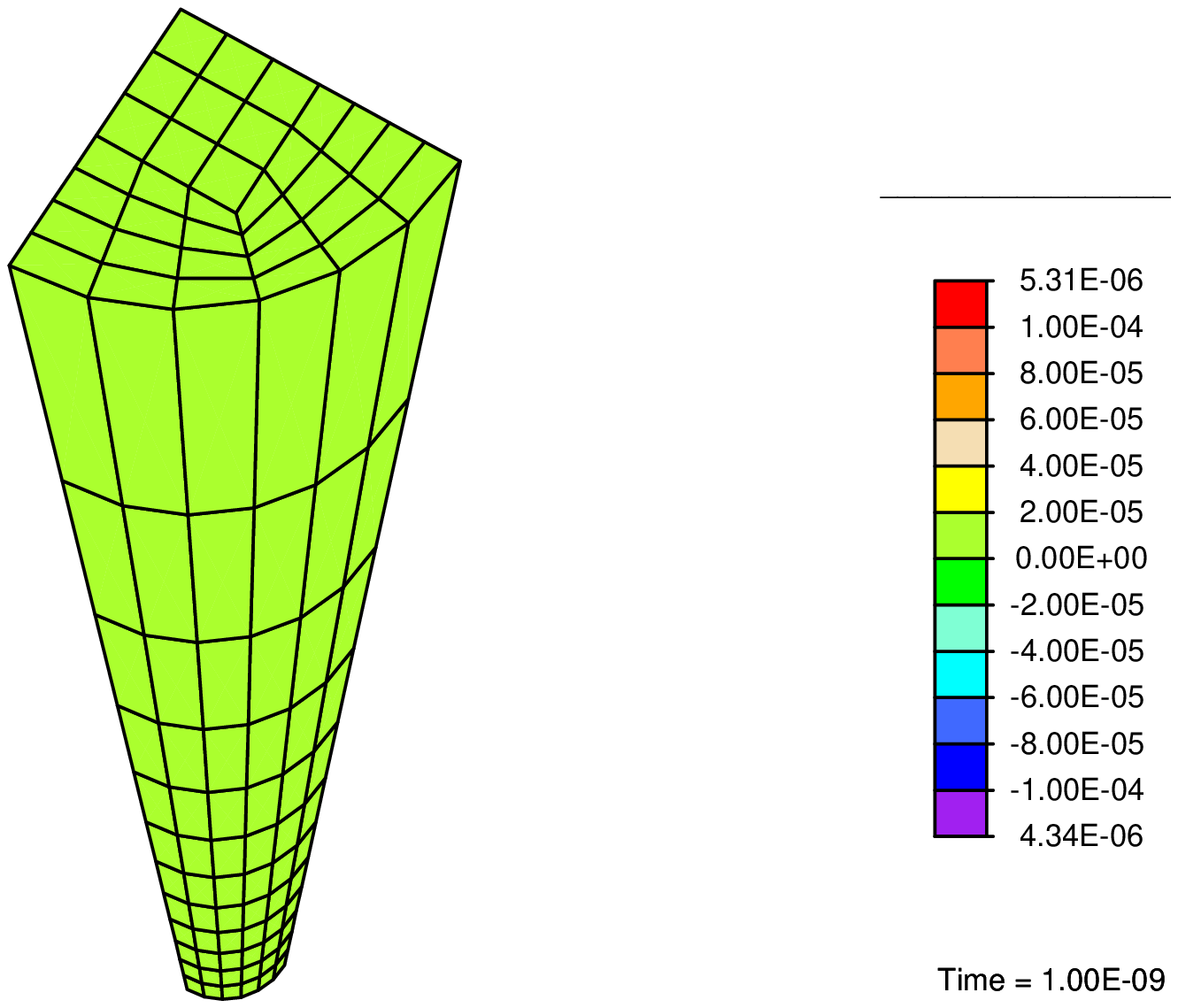}} \hskip 3cm (a)
\end{minipage}
\begin{minipage}[t]{7.5cm}
{\includegraphics[width=7.5cm]{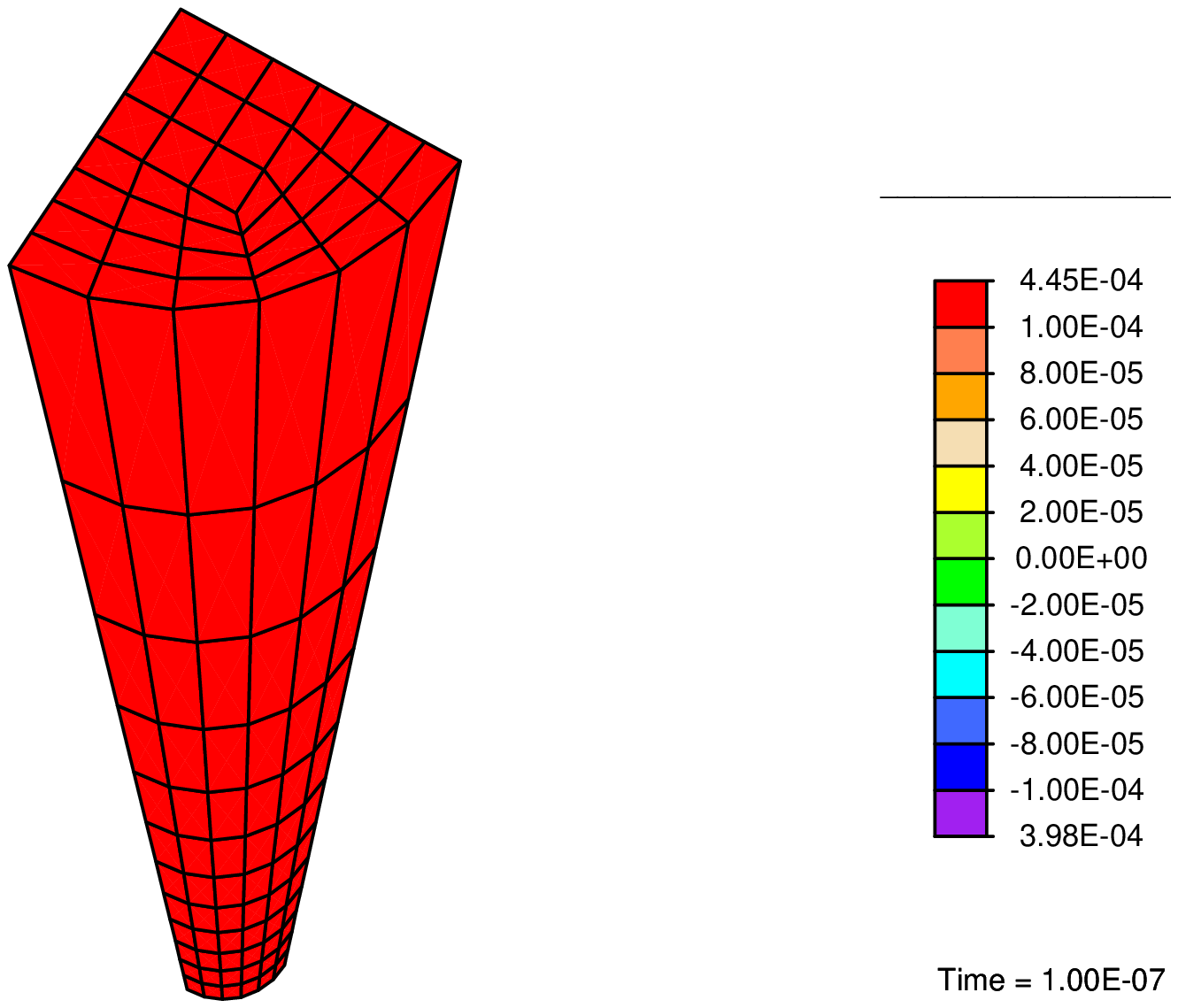}} \hskip 3cm (b)
\end{minipage}
\caption{Stress gradient-driven flux
($\mathrm{kg.m}^{-2}\mathrm{sec}$) in the $\be_3$ direction at $1
\,\mathrm{nanosec.}$ and $100\,\mathrm{nanosec.}$ after the
beginning of loading. The positive values indicate an upward flux
corresponding to a tensile $\sigma_{33}$ wave travelling
downwards.} \label{M1fig}
\end{figure}

\begin{figure}[ht]
\begin{minipage}[t]{7.5cm}
{\includegraphics[width=7.5cm]{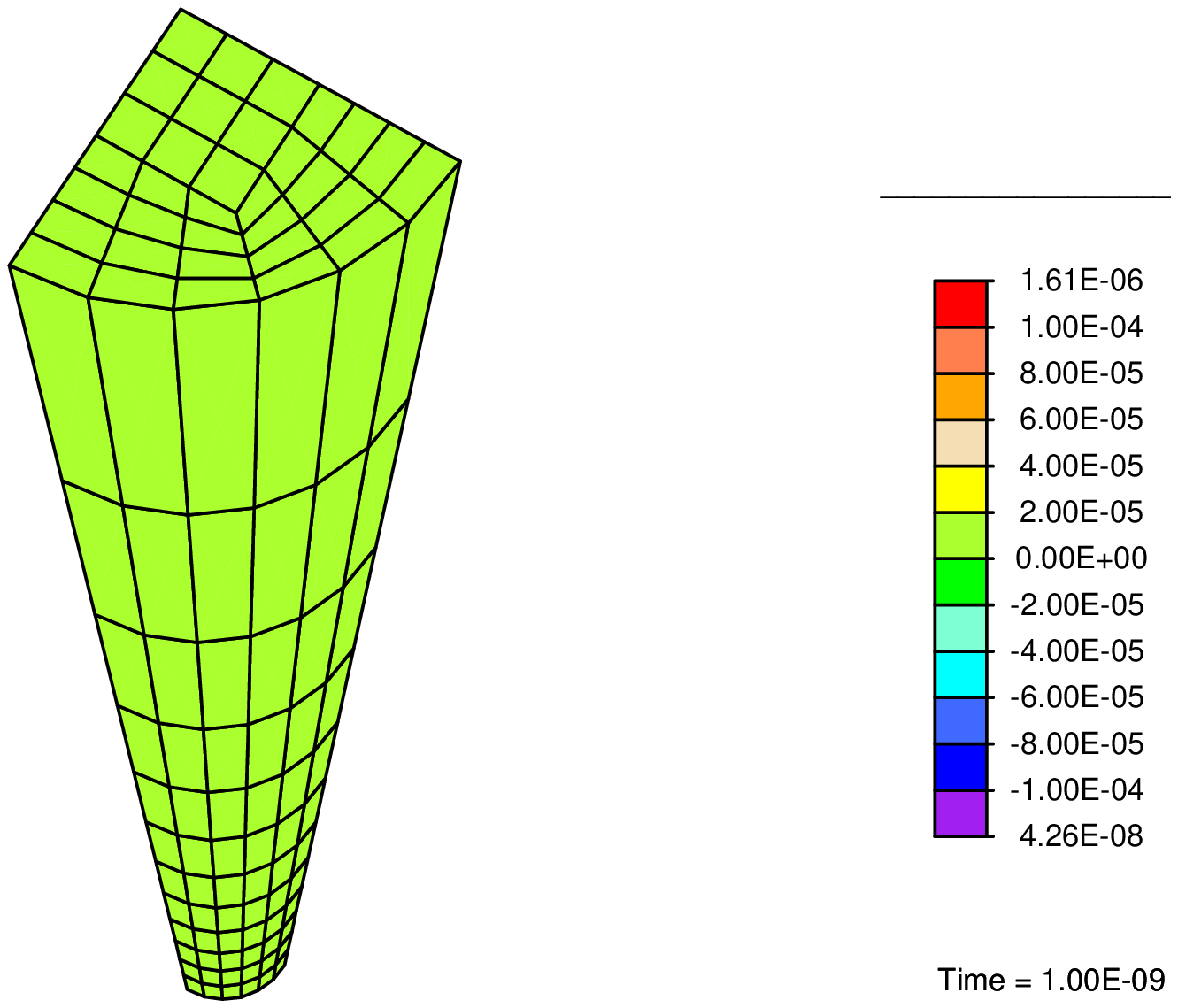}} \hskip 3cm (a)
\end{minipage}
\begin{minipage}[t]{7.5cm}
{\includegraphics[width=7.5cm]{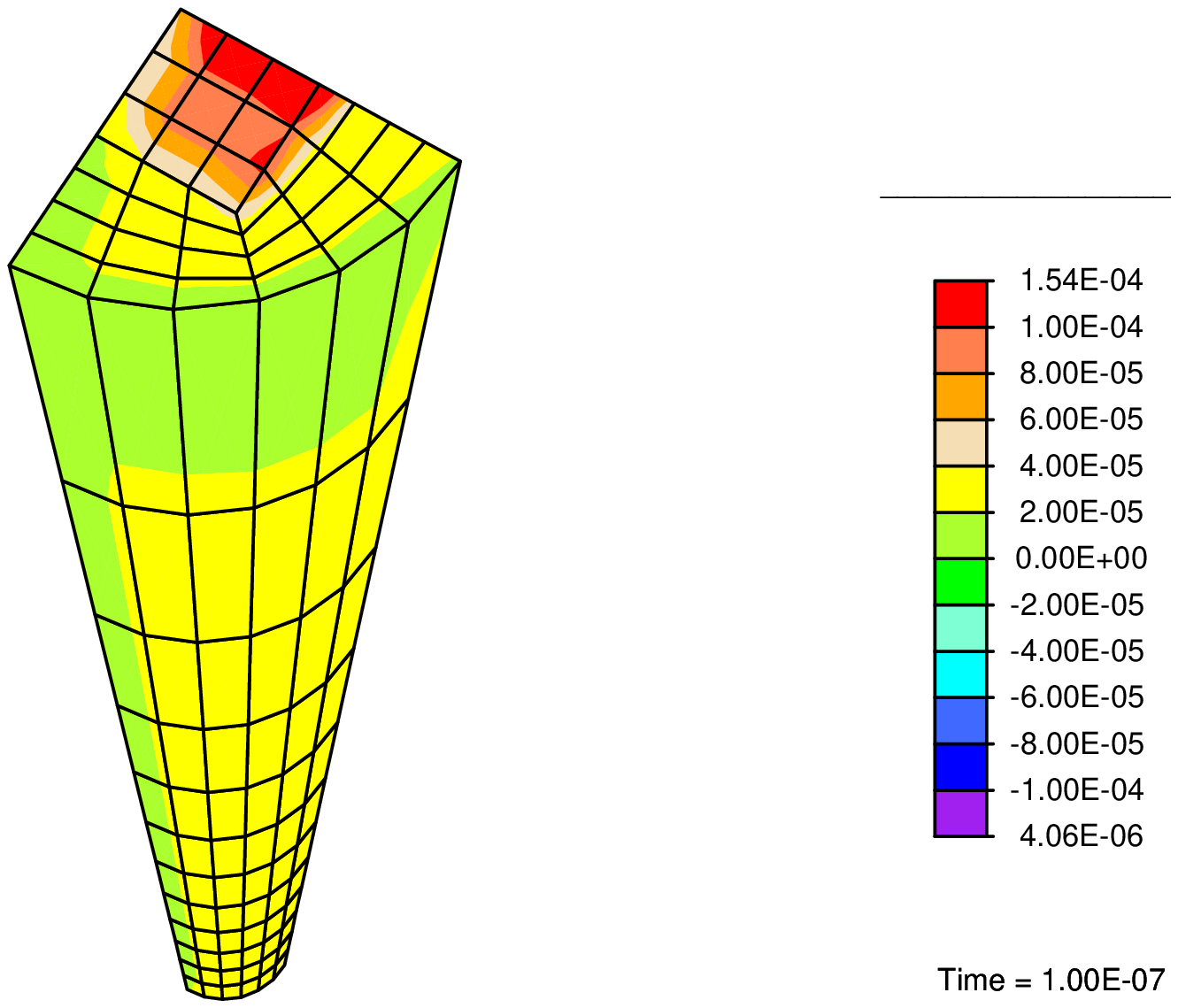}} \hskip 3cm (b)
\end{minipage}
\caption{Internal energy gradient-driven flux,
($\mathrm{kg.m}^{-2}\mathrm{sec}$) in the $\be_3$ direction at $1
\,\mathrm{nanosec.}$ and $100\,\mathrm{nanosec.}$ after the
beginning of loading. The positive values indicate an upward flux.
This corresponds to a lower energy near the top of the cylinder as
the tensile stress ($\sigma_{33}$) wave travels downward and
relaxes some of the strain energy of contraction.} \label{M2fig}
\end{figure}

\begin{figure}[ht]
\begin{minipage}[t]{7.5cm}
{\includegraphics[width=7.5cm]{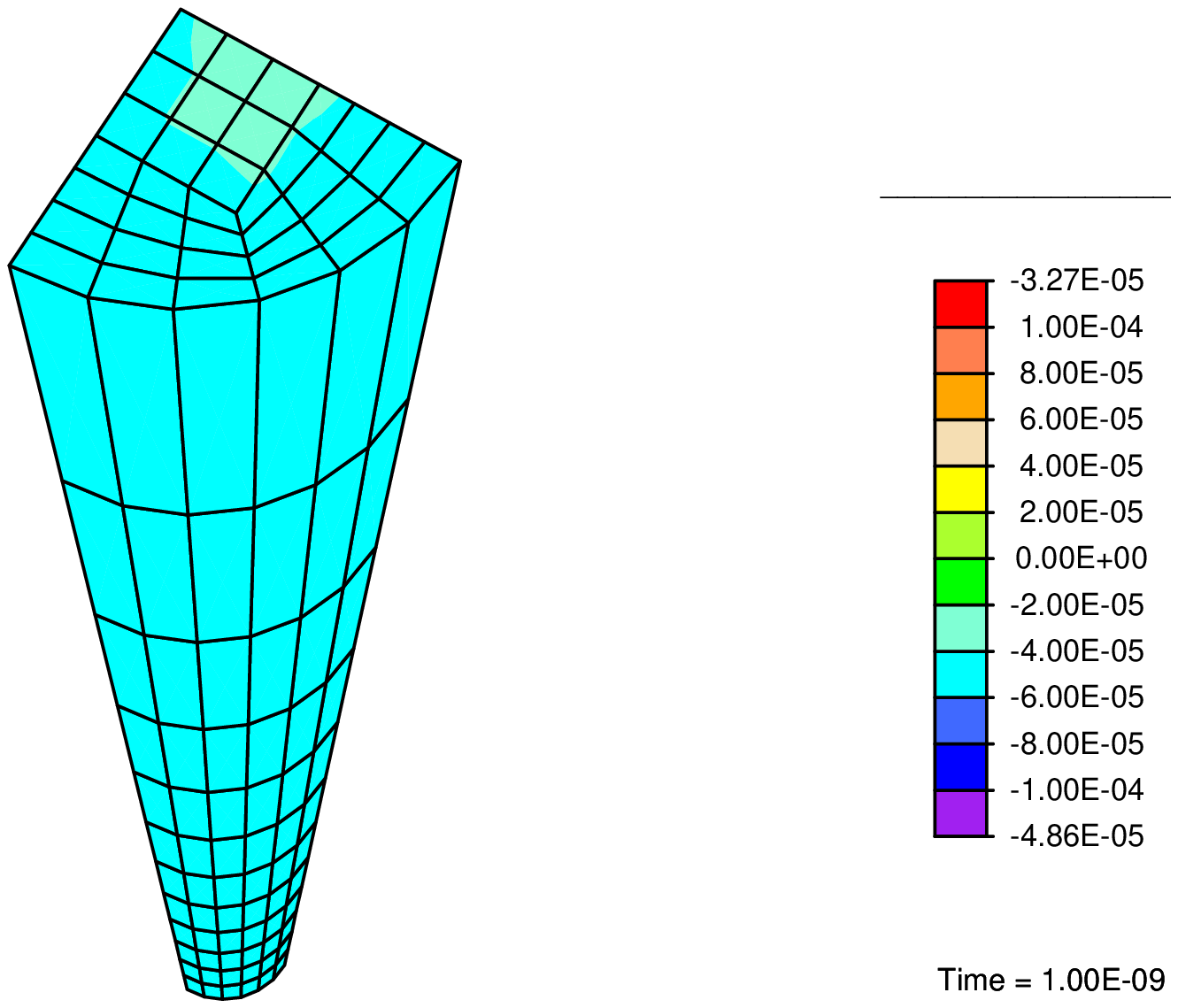}} \hskip 3cm (a)
\end{minipage}
\begin{minipage}[t]{7.5cm}
{\includegraphics[width=7.5cm]{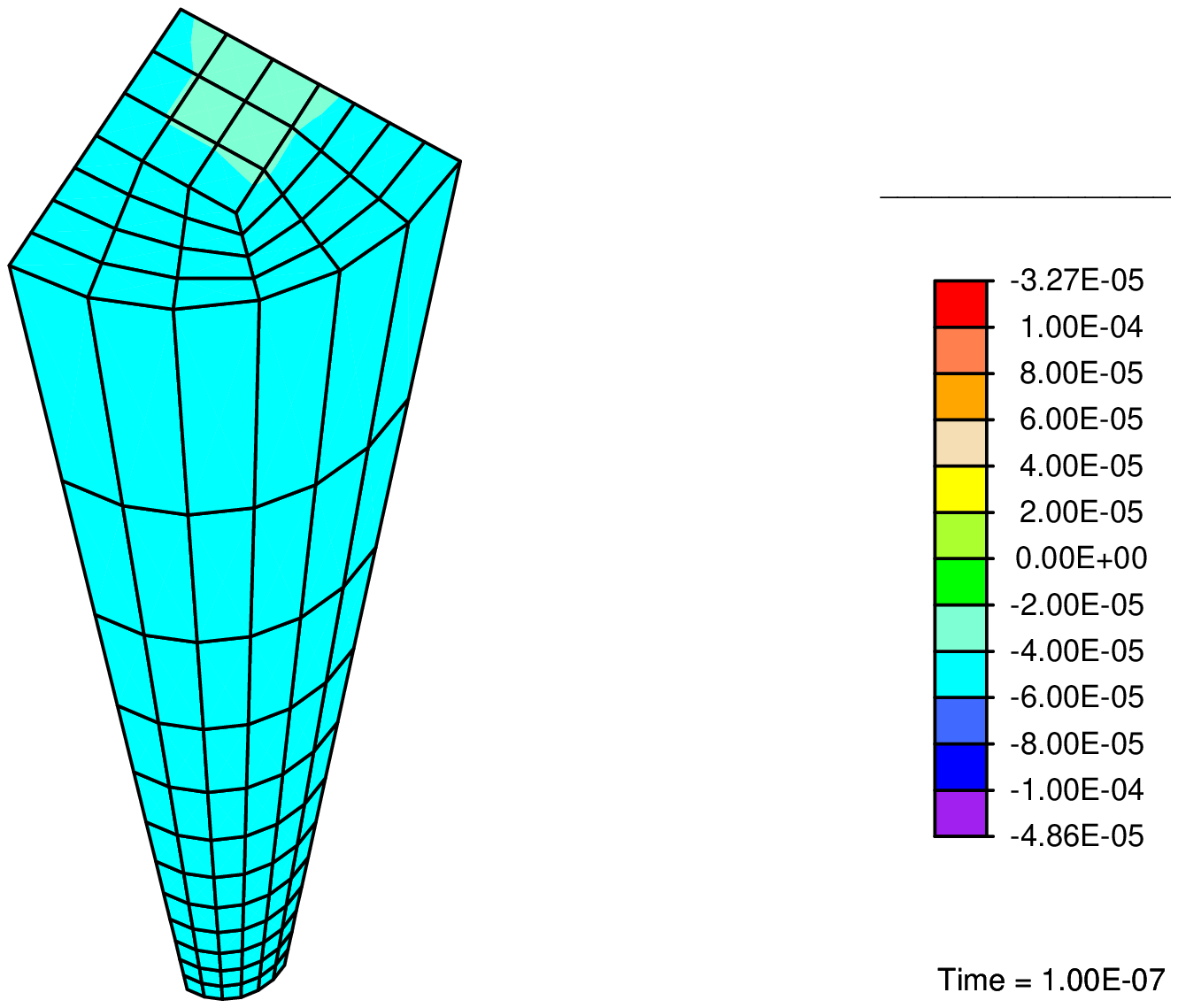}} \hskip 3cm (b)
\end{minipage}
\caption{Gravity-driven flux ($\mathrm{kg.m}^{-2}\mathrm{sec}$) in
the $\be_3$ direction at $1 \,\mathrm{nanosec.}$ and
$100\,\mathrm{nanosec.}$ after the beginning of loading. The
negative values indicate a downward flux, due to the action of
gravity.} \label{M3fig}
\end{figure}

\begin{figure}[ht]
\begin{minipage}[t]{7.5cm}
{\includegraphics[width=7.5cm]{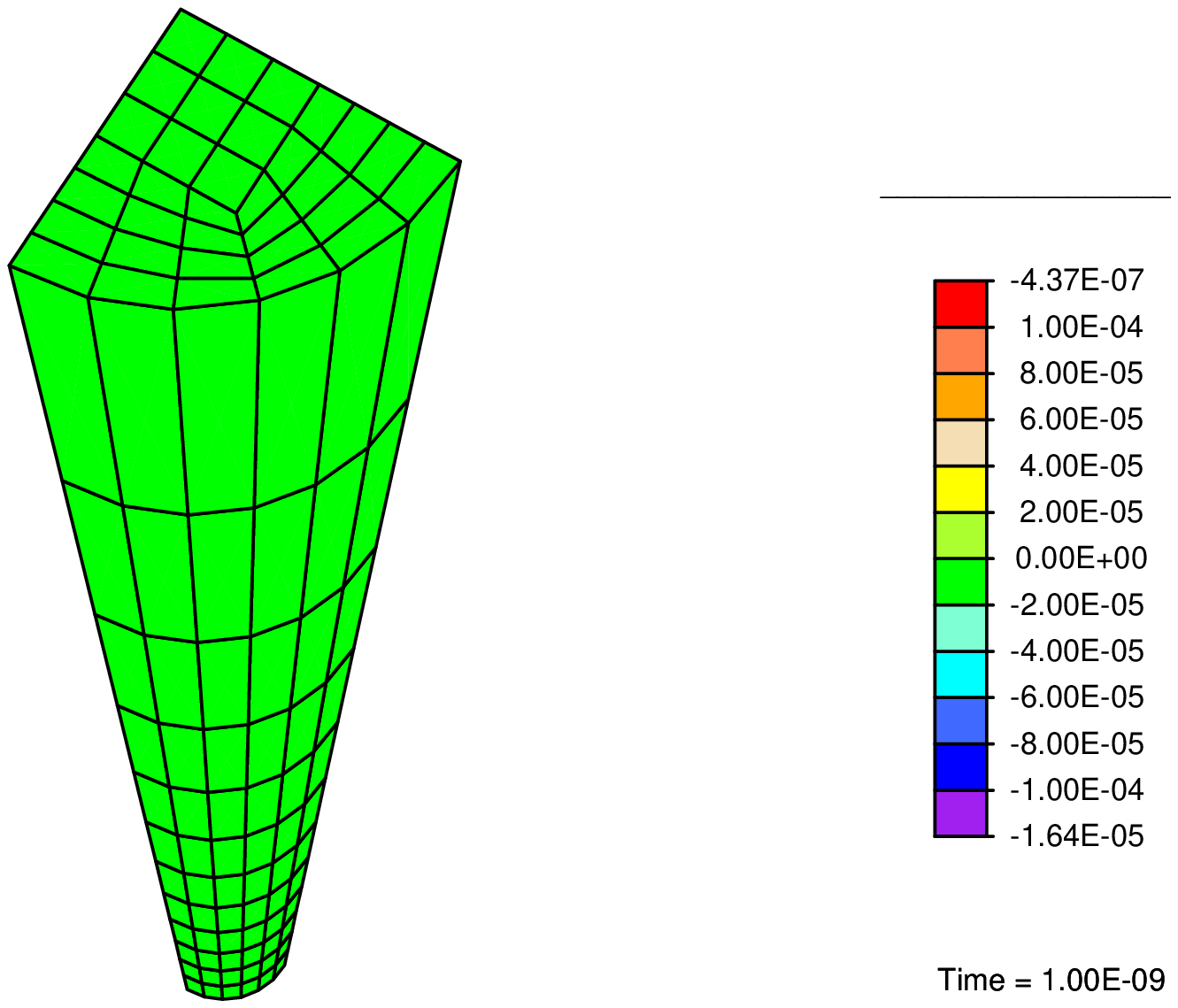}} \hskip 3cm (a)
\end{minipage}
\begin{minipage}[t]{7.5cm}
{\includegraphics[width=7.5cm]{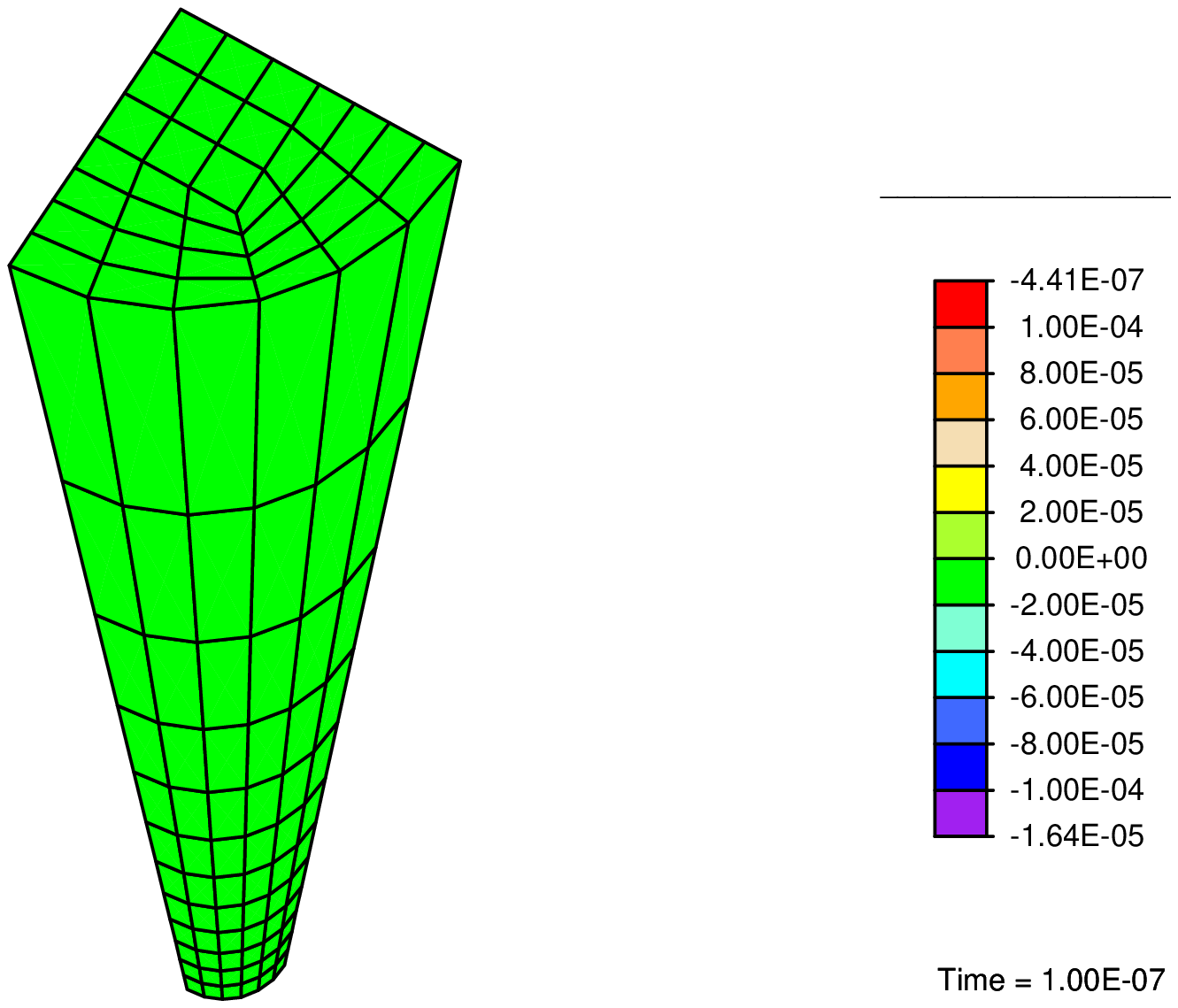}} \hskip 3cm (b)
\end{minipage}
\caption{Inertia-driven flux ($\mathrm{kg.m}^{-2}\mathrm{sec}$) in
the $\be_3$ direction at $1 \,\mathrm{nanosec.}$ and
$100\,\mathrm{nanosec.}$ after the beginning of loading. The
negative values indicate a downward flux as the tissue accelerates
upward.} \label{M4fig}
\end{figure}

\begin{figure}[ht]
\begin{minipage}[t]{7.5cm}
{\includegraphics[width=7.5cm]{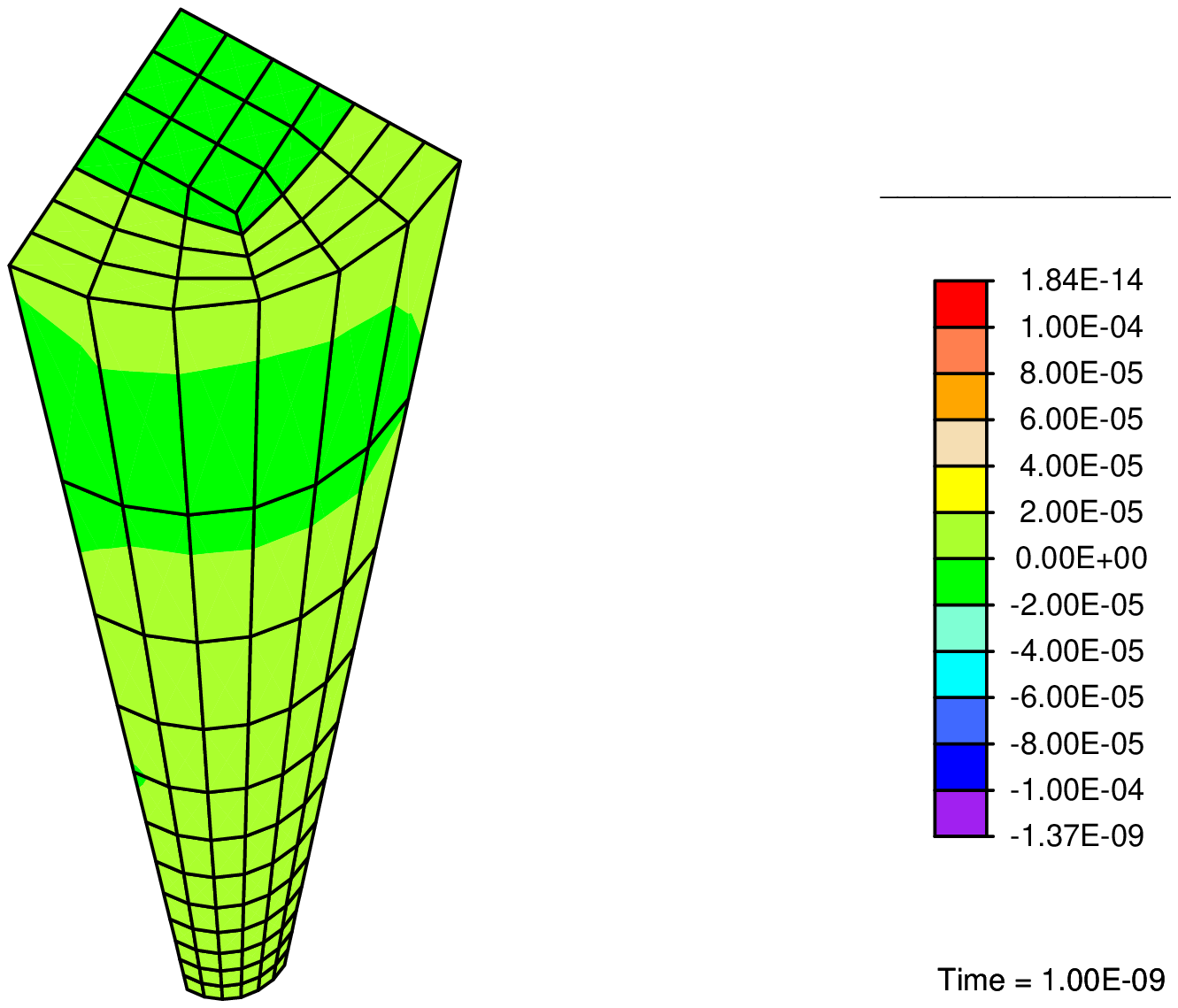}} \hskip 3cm (a)
\end{minipage}
\begin{minipage}[t]{7.5cm}
{\includegraphics[width=7.5cm]{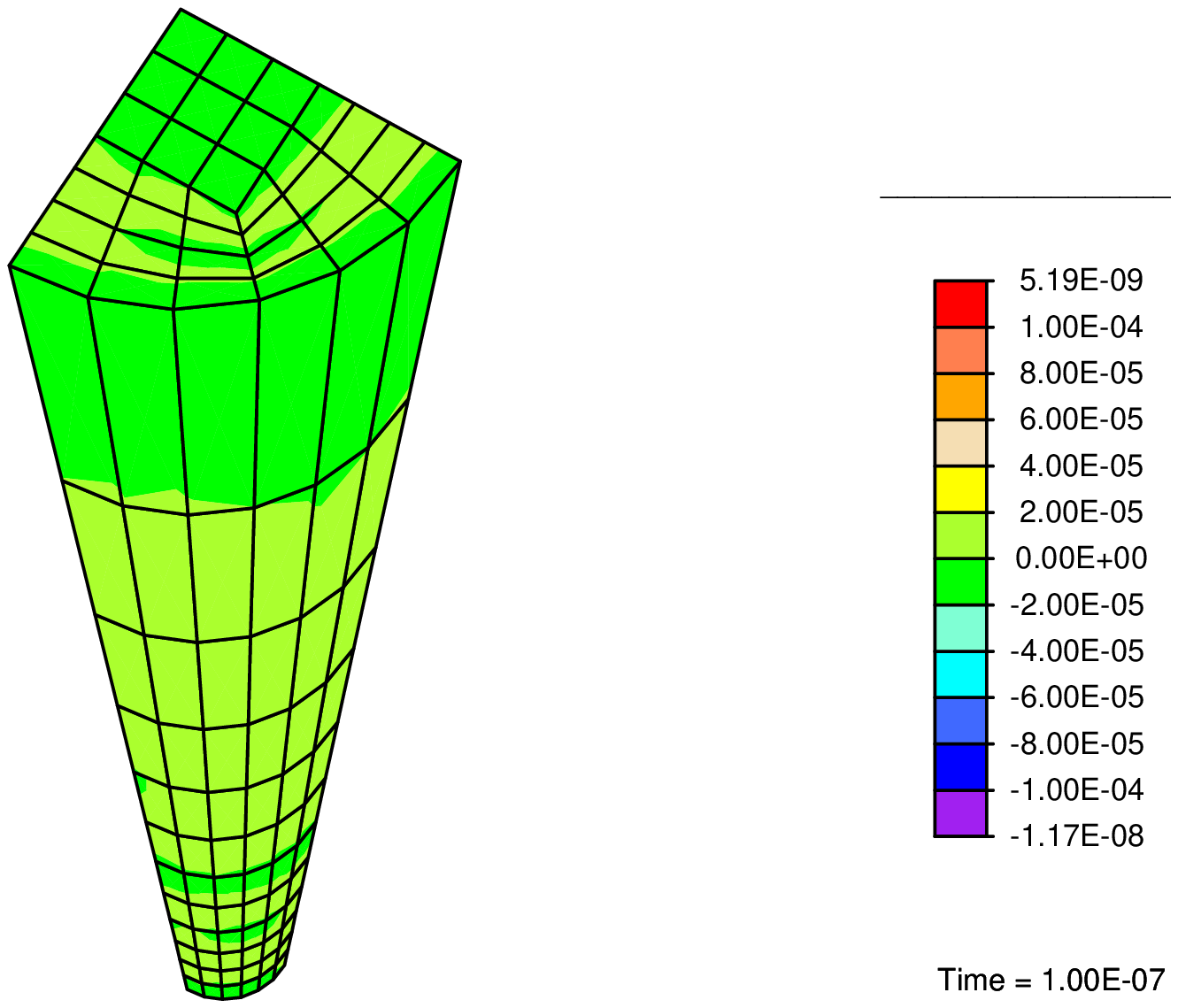}} \hskip 3cm (b)
\end{minipage}
\caption{Concentration gradient-driven flux
($\mathrm{kg.m}^{-2}\mathrm{sec}$) in the $\be_3$ direction at $1
\,\mathrm{nanosec.}$ and $100\,\mathrm{nanosec.}$ after the
beginning of loading. Note that the maximum and minimum values are
many orders of magnitude lower than for the other flux
contributions reported above. This is a demonstration of mechanics
influences dominating diffusion over the classical concentration
gradient contribution.} \label{M5fig}
\end{figure}

\begin{figure}[ht]
\begin{minipage}[t]{7.5cm}
{\includegraphics[width=7.5cm]{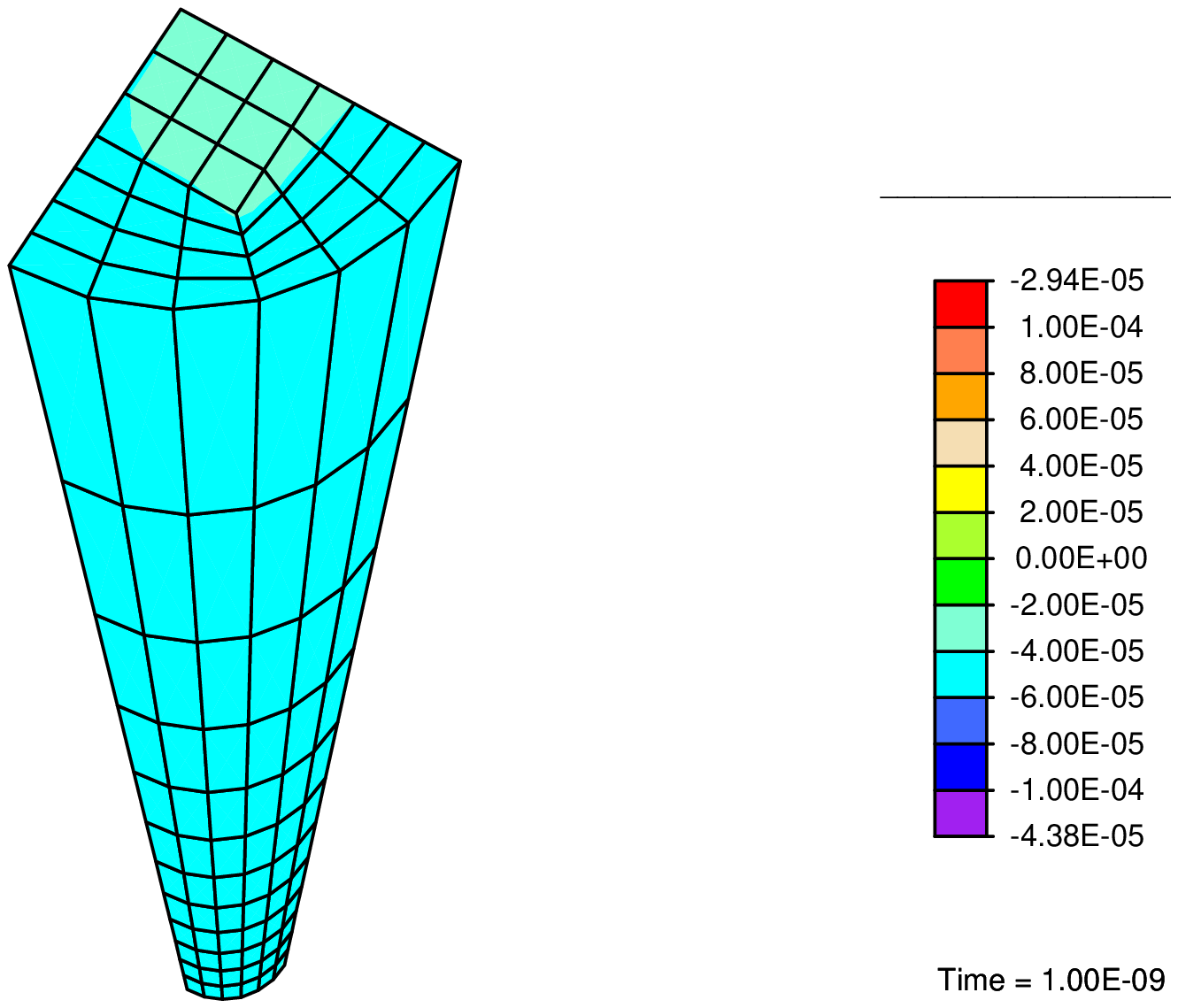}} \hskip 3cm (a)
\end{minipage}
\begin{minipage}[t]{7.5cm}
{\includegraphics[width=7.5cm]{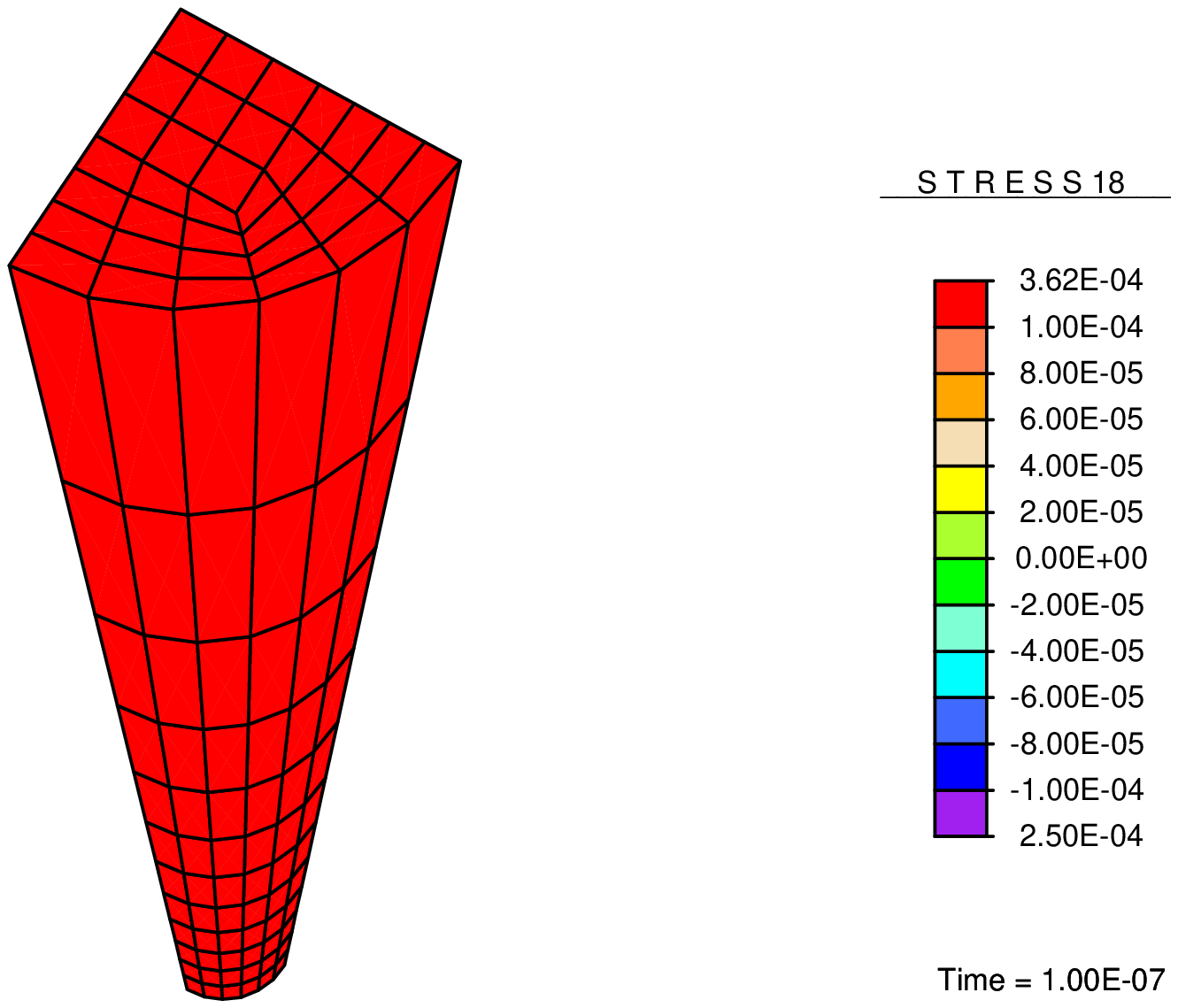}} \hskip 3cm (b)
\end{minipage}
\caption{Total flux ($\mathrm{kg.m}^{-2}\mathrm{sec}$) in the
$\be_3$ direction at $1 \,\mathrm{nanosec.}$ and
$100\,\mathrm{nanosec.}$ after the beginning of loading. The
positive values indicate an upward flux, dominated by the stress
gradient driven contribution.} \label{Mfig}
\end{figure}

\begin{figure}[ht]
\begin{minipage}[t]{7.5cm}
{\includegraphics[width=7.5cm]{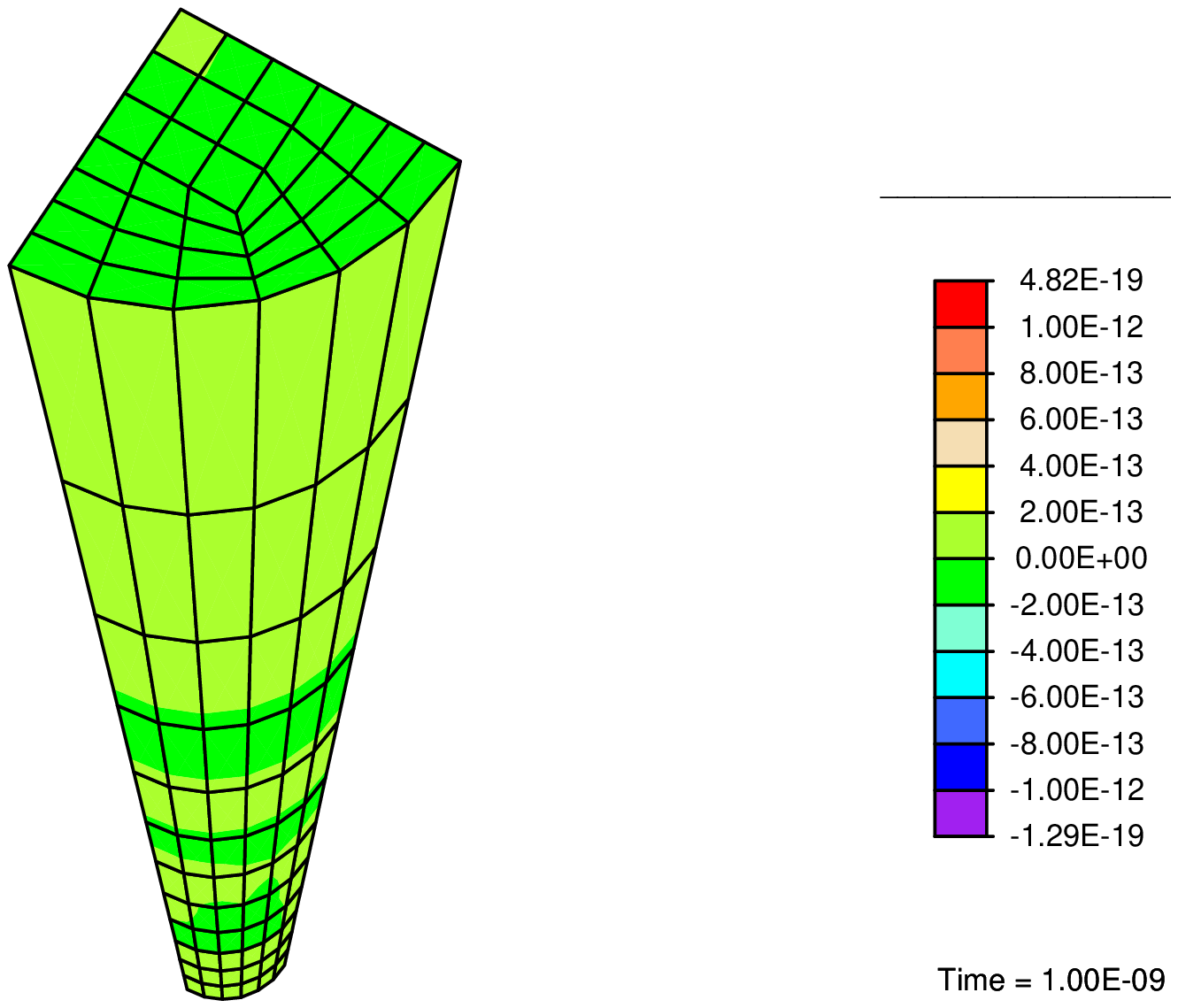}} \hskip 3cm (a)
\end{minipage}
\begin{minipage}[t]{7.5cm}
{\includegraphics[width=7.5cm]{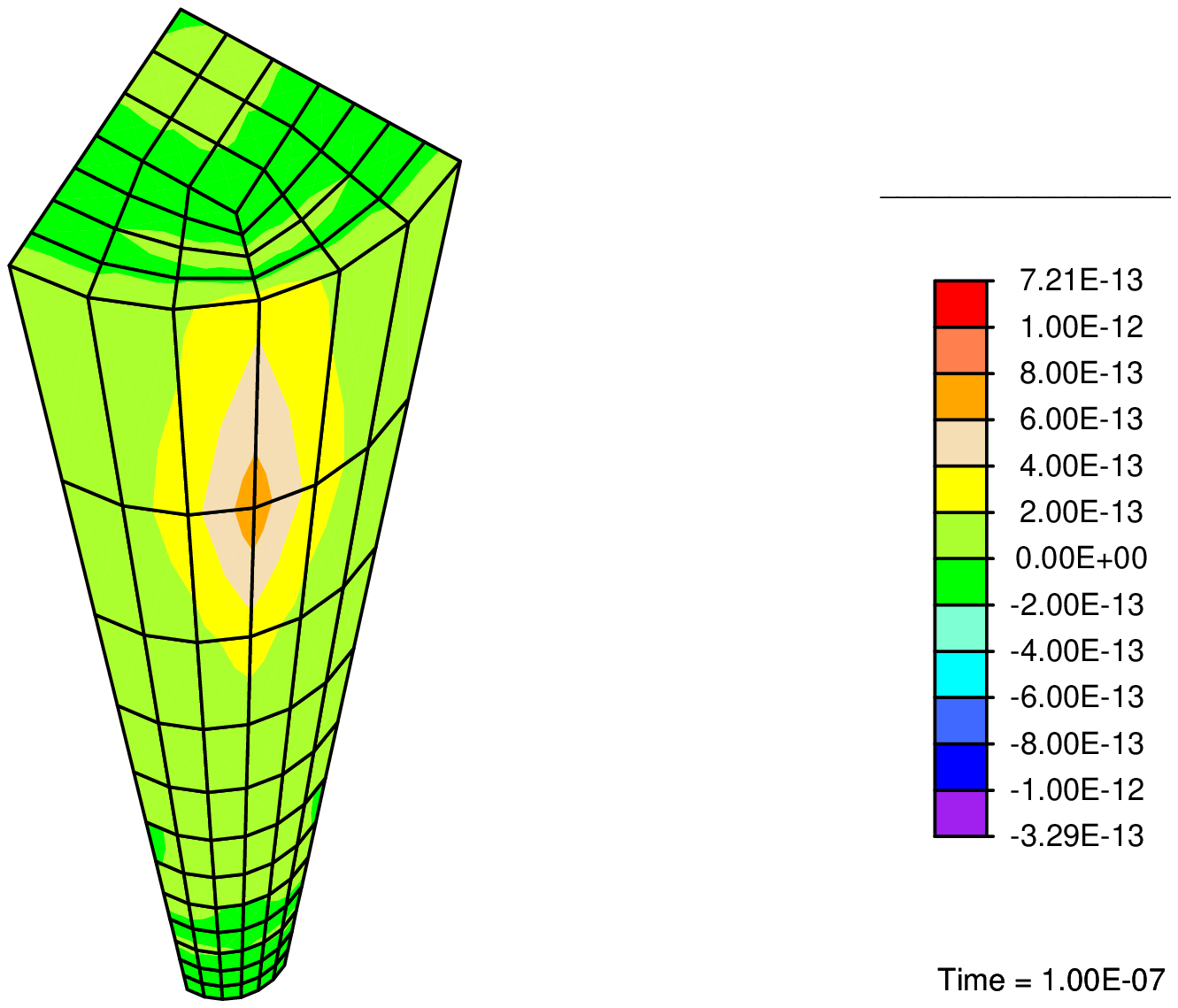}} \hskip 3cm (b)
\end{minipage}
\caption{Rate of fluid production, $\Pi^\mathrm{f}$
($\mathrm{kg.m}^{-3}.\mathrm{sec}^{-1}$), at $1
\,\mathrm{nanosec.}$ and $100\,\mathrm{nanosec.}$ after the
beginning of loading. The positive values indicate that the local
fluid concentrations have fallen below their initial values.}
\label{Pifig}
\end{figure}

The flux contributions in Figures \ref{M1fig}--\ref{Mfig} can be
summarized as follows: The fluid flux is dominated by the
contribution from the stress gradient in the $\be_3$ direction.
The latter arises as the stress ($\sigma_{33}$) wave of tension
travels down the cylinder in the first few microseconds after
application of the load (the time taken to travel the length of
the cylinder is $12 \,\mu\mathrm{sec}$). Additionally, as the
fluid concentration changes due to the flux, it causes a further
change in the stress (Section \ref{sect5}). Other flux terms are
qualitatively sensible; i.e., their directions are consistent with
the physics of the problem, as argued in each of the figure
captions\footnote{In order to compare the flux contributions, they
have all been plotted on the same scale: $-1\times
10^{-4}--1\times 10^{-4} \,\mathrm{kg.m}^{-2}\mathrm{sec}$.
However the plots also show the maximum and minimum field values
at the top and bottom of the legend bars. These values represent a
better comparison of the relative flux magnitudes.}. There is some
loss of axial symmetry in a few of the plots due to the coarseness
of the finite element mesh for this example. It appears that
spatial oscillations in the solution lead to a further loss of
symmetry in Figures \ref{M2fig}b--\ref{M3fig}b and \ref{Mfig}a.
These oscillations arise due to large and dominant advective
terms, and need to be remedied by stabilized finite element
methods. Here, we only aim to demonstrate that various driving
forces for mass transport are in agreement with their theoretical
underpinnings in the paper. The resorption of the solid phase is
shown indirectly in Figure \ref{Pifig}. A positive fluid source,
$\Pi^\mathrm{f}$, means that $\Pi^\mathrm{s} < 0$. Since
$\Pi^\mathrm{s}$ is the only term balancing
$\partial\rho_0^\mathrm{s}/\partial t$ [see (\ref{massballocA})],
it follows that $\partial\rho_0^\mathrm{s}/\partial t < 0$.

\section{Discussion and conclusions}
\label{sect7}

A general framework for growth of biological tissue has been
presented in this paper. While simplified models were used for
source terms, $(\Pi^\iota)$, they can be formulated on the basis
of the kinetics of chemical reactions in order to develop more
realistic growth laws. This approach, we believe, is fundamental
to a proper treatment of mass transport in tissue. Results are
obtained that differ fundamentally from the classical setting of
continuum mechanics. Most notable among these differences are the
mass fluxes driven by gradients in stress, strain, energy and
entropy, in addition to body force and inertia. Importantly,
though our treatment differs from classical mixture theory, the
two are fully consistent as established at several points in this
paper. The balance laws in Section \ref{sect2} and \ref{sect3}
introduce a degree of coupling between the phenomena of mass
transport and mechanics. This is visible most transparently in the
balance of linear momentum (\ref{ballinmomrefI}) that includes
mass fluxes, $\bM^\iota$. The balance of mass, described by
Equations (\ref{massballocA}) and (\ref{massballocI}), is also
dependent upon the mechanics as the discussion in Section
\ref{sect5.1} makes clear. Notably, this ensures
mechanics-mediated mass transport even with a mass source that is
independent of strain/stress, as the discussion at the end of
Section \ref{sect5.3} establishes. The discussion in Sections
\ref{sect5.1}--\ref{sect5.3} provides many insights into the
nature of this coupling. The mechanics problem also has a
constitutive dependence upon mass concentration, via
(\ref{stress-constrelI}) and the fact that the growth deformation
gradient tensor, $\bF^{\mathrm{g}^\iota}$ is determined by the
concentration. The viscoelastic nature of the composite tissue
would emerge naturally from a model incorporating a hyperelastic
solid and viscous fluid.

We have formally allowed all species to be load bearing and
develop a stress. At the scales that are of interest in a tissue,
the only relevant load bearing species are the solid and fluid
phases. Nevertheless, inasmuch as a transported species such as a
nutrient has a molecular structure that can be subject to loads at
the scale of pico-newtons, it is not inconsistent to speak of the
partial stress of this species. Since the constitutive relation
(\ref{stress-constrelI}) indicates that the partial stresses are
scaled by concentrations, the contribution to total stress from
any species besides the solid and fluid phases will be negligible.

We have chosen to leave remodelling out of our formulation in this
communication, to focus upon the above issues. Remodelling
includes any evolution in properties, state of stress, material
symmetry, volume or shape brought about by microstructural
changes. In the development of biological tissue, growth and
remodelling occur simultaneously. As density changes due to
growth, the material also remodels by microstructural evolution
within the neighborhood of each point. A rigorous treatment of
this phenomenon has been presented in the continuum mechanical
setting in a companion paper \citep{remodelpaper}.

% main text
\bibliography{mybib}
\bibliographystyle{elsart-harv}
\end{document}